\documentclass[twocolumn,tighten]{aastex701}

\usepackage{amsmath}

\defcitealias{4FGL}{4FGL}
\defcitealias{4FGL-DR4}{4FGL-DR4}
\defcitealias{2FIG}{2FIG}
\defcitealias{3PC}{3PC}
\defcitealias{PaperI}{Paper I}
\defcitealias{PaperII}{Paper II}

\shorttitle{\textit{Einstein@Home} Inner Galaxy Pulsar Searches}
\shortauthors{Clark, C.~J., et al.}

\begin{document}

\title{\textit{Einstein@Home} Searches for Gamma-ray Pulsars in the Inner Galaxy}

\author[0000-0003-4355-3572]{C.~J.~Clark}
\email[show]{colin.clark@aei.mpg.de}
\affiliation{Max Planck Institute for Gravitational Physics (Albert Einstein Institute), D-30167 Hannover, Germany}
\affiliation{Leibniz Universit\"at Hannover, D-30167 Hannover, Germany}

\author[0000-0003-2759-5625]{M.~Di~Mauro}
\email{placeholder}
\affiliation{Istituto Nazionale di Fisica Nucleare, Sezione di Torino, I-10125 Torino, Italy}

\author[0000-0003-3536-4368]{J.~Wu}
\email{placeholder}
\affiliation{Max-Planck-Institut f\"ur Radioastronomie, Auf dem H\"ugel 69, D-53121 Bonn, Germany}

\author[0000-0003-4285-6256]{B.~Allen}
\email{placeholder}
\affiliation{Max Planck Institute for Gravitational Physics (Albert Einstein Institute), D-30167 Hannover, Germany}
\affiliation{Leibniz Universit\"at Hannover, D-30167 Hannover, Germany}

\author[0000-0003-0679-8562]{O.~Behnke}
\email{placeholder}
\affiliation{Max Planck Institute for Gravitational Physics (Albert Einstein Institute), D-30167 Hannover, Germany}
\affiliation{Leibniz Universit\"at Hannover, D-30167 Hannover, Germany}

\author[0000-0001-5296-7035]{H.~B.~Eggenstein}
\email{placeholder}
\affiliation{Max Planck Institute for Gravitational Physics (Albert Einstein Institute), D-30167 Hannover, Germany}
\affiliation{Leibniz Universit\"at Hannover, D-30167 Hannover, Germany}

\author[0000-0002-2332-0459]{B.~Machenschalk}
\email{placeholder}
\affiliation{Max Planck Institute for Gravitational Physics (Albert Einstein Institute), D-30167 Hannover, Germany}
\affiliation{Leibniz Universit\"at Hannover, D-30167 Hannover, Germany}

\author[0000-0002-5775-8977]{L.~Nieder}
\email{placeholder}
\affiliation{Max Planck Institute for Gravitational Physics (Albert Einstein Institute), D-30167 Hannover, Germany}
\affiliation{Leibniz Universit\"at Hannover, D-30167 Hannover, Germany}

\author[0000-0001-6566-1246]{P.~M.~Saz~Parkinson}
\email{placeholder}
\affiliation{Santa Cruz Institute for Particle Physics, Department of Physics and Department of Astronomy and Astrophysics, University of California at Santa Cruz, Santa Cruz, CA 95064, USA}

\author[0000-0002-8395-957X]{A.~Ashok}
\email{placeholder}
\affiliation{Max Planck Institute for Gravitational Physics (Albert Einstein Institute), D-30167 Hannover, Germany}
\affiliation{Leibniz Universit\"at Hannover, D-30167 Hannover, Germany}
\affiliation{Department of Physics, Oregon State University, Corvallis, OR 97331, USA}

\author[0000-0002-9032-7941]{P.~Bruel}
\email{placeholder}
\affiliation{Laboratoire Leprince-Ringuet, CNRS/IN2P3, \'Ecole polytechnique, Institut Polytechnique de Paris, 91120 Palaiseau, France}

\author[0009-0002-4068-7911]{B.~McGloughlin}
\email{placeholder}
\affiliation{Max Planck Institute for Gravitational Physics (Albert Einstein Institute), D-30167 Hannover, Germany}
\affiliation{Leibniz Universit\"at Hannover, D-30167 Hannover, Germany}

\author[0000-0002-1007-5298]{M.~A.~Papa}
\email{placeholder}
\affiliation{Max Planck Institute for Gravitational Physics (Albert Einstein Institute), D-30167 Hannover, Germany}
\affiliation{Leibniz Universit\"at Hannover, D-30167 Hannover, Germany}

\author[0000-0002-1873-3718]{F.~Camilo}
\email{placeholder}
\affiliation{South African Radio Astronomy Observatory, 2 Fir Street, Observatory, 7925, South Africa}

\author[0000-0002-0893-4073]{M.~Kerr}
\email{placeholder}
\affiliation{Space Science Division, Naval Research Laboratory, Washington, DC 20375-5352, USA}

\author[0000-0001-5624-4635]{P.~Voraganti~Padmanabh}
\email{placeholder}
\affiliation{Max Planck Institute for Gravitational Physics (Albert Einstein Institute), D-30167 Hannover, Germany}
\affiliation{Leibniz Universit\"at Hannover, D-30167 Hannover, Germany}

\author[0000-0001-5799-9714]{S.~M.~Ransom}
\email{placeholder}
\affiliation{National Radio Astronomy Observatory, 1003 Lopezville Road, Socorro, NM 87801, USA}
\correspondingauthor{Clark, C.~J.}

\begin{abstract}
  The \textit{Fermi} Large Area Telescope (LAT) has revealed a mysterious
  extended excess of GeV gamma-ray emission around the Galactic Center, which
  can potentially be explained by unresolved emission from a population of
  pulsars, particularly millisecond pulsars (MSPs), in the Galactic bulge. We
  used the distributed volunteer computing system \textit{Einstein@Home} to
  search the \textit{Fermi}-LAT data for gamma-ray pulsations from sources in
  the inner Galaxy, to try to identify the brightest members of this putative
  population. 
  We discovered four new pulsars, including one new MSP and one 
  young pulsar whose angular separation to the Galactic Center of 0.93$\degr$ is 
  the smallest of any known gamma-ray pulsar. We demonstrate 
  a phase-resolved difference imaging technique that allows the flux from this 
  pulsar to be disentangled from the diffuse Galactic Center emission. No radio pulsations were detected from the four new pulsars in archival radio observations or during the MPIfR-MeerKAT Galactic Plane Survey.  While the
  distances to these pulsars remain uncertain, we find that it is more likely
  that they are all foreground sources from the Galactic disk, rather than
  pulsars originating from the predicted bulge population. Nevertheless, our
  results are not incompatible with an MSP explanation for the GC excess, as only 
  one or two members of this population would have been detectable in our
  searches.
\end{abstract}

\section{Introduction}
Observations by the Large Area Telescope (LAT) onboard the \textit{Fermi
  Gamma-ray Space Telescope} \citep{LAT} have revealed an excess of GeV
gamma-ray emission in the inner region of our Galaxy (first discovered by
\citealt{Goodenough2009}, and recently reviewed by \citealt{MurgiaReview}).
While the presence of this excess has been confirmed in multiple independent
analyses
\citep[e.g.,][]{Hooper2011,Abazajian2012,Calore2015,FermiCollab+GCExcess,
  DiMauro2021}, which generally find the excess to be spatially extended and to have a
spectrum that peaks in the GeV range, the large and uncertain fore/background
flux from the interstellar medium and nearby Galactic sources makes it
challenging to robustly measure its spectral and spatial properties. As a
result, the origin of this excess remains unclear.

Two main hypotheses for the origin of this Galactic Center (GC) excess have
emerged. The first is that it is the signature of a central concentration of
dark matter, and is produced through the annihilation of weakly-interacting
massive particles (WIMPs). This hypothesis is based on the observations that the
spatial distribution of the excess appears to be consistent with the generalized
Navarro-Frenk-White \citep[NFW,][]{NFW} distribution expected for dark matter,
and that the shape of the spectrum can be reproduced by the annihilation of
WIMPs with a mass in the plausible range of 30--300 GeV \citep[][and references
  therein]{MurgiaReview}.

The leading alternative to this dark-matter explanation is that the GeV excess
is due to the combined emission from an unresolved, centrally concentrated population 
of millisecond pulsars (MSPs)
numbering in the thousands or tens of thousands \citep[e.g.][]{Abazajian2011+GCMSPs, Calore2016, Gonthier2018+GCMSPs, Ploeg2020}. 
It has long been predicted
that such a ``bulge'' population of MSPs might exist in the inner Galaxy
\citep[e.g.][]{Wang2005,Wharton2012,Macquart2015}, where the larger stellar density and encounter rate
encourages binary MSP formation, and where the remnants of disrupted MSP-rich
globular clusters might be expected to reside \citep{Tremaine1975}. Further
evidence for this population comes from the increased density of low-mass X-ray
binaries (the progenitors of MSPs) at the GC \citep{Muno2005} as well as
at the center of the Andromeda galaxy \citep{Abazajian2012}. The \textit{Fermi}
LAT has detected pulsed gamma-ray emission from more than 150 MSPs \citep[][hereafter \citetalias{3PC}]{3PC}, and these objects have
characteristic curved gamma-ray spectra peaking at GeV energies,
similar to that of the GC excess.

Confirming or ruling out either of these hypotheses is a crucial, although extremely
challenging goal. It has not yet proved possible to distinguish between these
hypotheses in the gamma-ray data alone. Both dark-matter and MSP models can
reproduce the observed properties of the GC excess, within their large
systematic uncertainties, and the gamma-ray spectra predicted by these models
only differ significantly at low gamma-ray energies, where there are high levels
of source/background confusion, and at high energies where photon counting statistics 
are lower (although see \citealt{Manconi2024}).

Additional observations, of other sources with similar gamma-ray emission or at other wavelengths, 
will likely be required to settle the debate. In the case of the dark-matter hypothesis,
evidence would come from additional astrophysical detections, e.g. the discovery
of similar emission from the Milky Way's satellite dwarf spheroidal galaxies
\citep{FermiCollab2011+DwarfGalaxies}, or from the direct detection of dark
matter particles with similar masses and cross-sections in the laboratory.

For the MSP hypothesis, searches with radio telescopes could detect individual
members of this predicted ``bulge'' population, and with large enough numbers
this could be distinguished from the known population in the Galactic
disk. However, discovering new MSPs in this region of the Galaxy is particularly
challenging. First, MSPs in the inner Galaxy are distant, and therefore faint;
more than 97\% of Galactic MSPs (excluding those in globular clusters)
in the ATNF Pulsar Catalogue \citep{ATNF}\footnote{\url{https://www.atnf.csiro.au/research/pulsar/psrcat}} are closer than the 8.2~kpc distance to the GC. Second,
the dense interstellar medium in the inner Galaxy, and the resulting large
column density of free electrons along that line-of-sight, means that radio emission
from MSPs is scattered (spread out in time) and
dispersed (delayed in time with a frequency dependence), with both of these
effects making their pulsations harder to detect. Likely as a result of these
effects, only one MSP has been discovered within 1$^{\circ}$ of the GC
\citep{Lower2024}, and simulations by \citet{Macquart2015} and
\citet{Calore2016} predict that only a small number of MSPs from this putative
bulge population may be found in the most sensitive surveys currently possible
with MeerKAT\footnote{While the FAST telescope is more sensitive, its location
in the Northern Hemisphere means that it cannot observe the Galactic
Center.}. Discovering new MSPs in the inner Galaxy is a central goal for the
MPIfR-MeerKAT Galactic Plane Survey \citep[MMGPS,][]{MMGPS}, which will
search the GC at S-band ($\sim2$--$3$~GHz), where the higher radio frequency
helps mitigate some of the deleterious propagation effects, but it is possible
that next-generation telescopes e.g. the SKA, ngVLA and DSA-2000 will be
required to properly probe this predicted population \citep{Calore2016}.
 
However, the MSP hypothesis has one further property that may enable some
limited tests in the gamma-ray data itself, which is that the gamma-ray emission
from individual MSPs is pulsed in time. Detecting an individual gamma-ray pulsar
in the \textit{Fermi}-LAT point source at the position of the GC, 4FGL~J1745.6-2859 
\citep{4FGL}, is unlikely: under this hypothesis, this point source represents the combined 
emission of hundreds or even thousands of pulsars, and so a single pulsar 
would have to be extremely bright to stand out on top of this. 
The GeV excess is, however, extended, possibly up to around $10\degr$ from the GC
\citep{FermiCollab2011+DwarfGalaxies,DiMauro2021}, and we may therefore hope
that there are rare, bright members of this population lying $\gtrsim 1^{\degr}$ 
from the GC, where they can be individually resolved above the local background flux. 
Gamma-ray pulsations have now been detected from
individual MSPs in four globular clusters
\citep{Freire2011+NGC6624A,Wu2013+M28A,Zhang2022+NGC6652B,Zhang2023+NGC6341A},
where a similar (albeit much smaller) background from the combined emission of
other pulsars in the cluster is (likely) present. \citet{Malyshev2024} also find
that a fraction of the extended GeV excess has now been resolved into individual
point sources in the latest \textit{Fermi}-LAT data, as would be expected for a
population of discrete sources.

Detecting gamma-ray pulsations from a previously unknown pulsar requires
computationally-demanding searches over at least four parameters (the spin
period, spin-down rate and two sky-position parameters). To make these searches
computationally feasible, necessary compromises must be made which reduce their
sensitivity \citep{Atwood2006}, meaning that only fairly bright isolated pulsars can be
discovered in this route (we quantify this in Section~\ref{s:discussion}). 
Nevertheless, such searches have discovered more than
80 new pulsars \citep[e.g.,][]{Abdo2009,SazParkinson2010,Pletsch2012+9PSRs,3PC},
including 7 new gamma-ray MSPs
\citep{Clark2018+EAHMSPs,3PC}. Three binary MSPs have also been found using the 
same methods \citep{Pletsch2012+J1311,Nieder2020+J1653,Clark2021+J2039}, 
but these required precise constraints on their orbital parameters and sky positions through optical 
observations of their binary companions to account for their orbital Doppler shifts. 
Binary MSPs cannot yet be discovered 
in these searches without prior orbital constraints as the orbital parameter volume is prohibitively large.
While most MSPs in the putative bulge 
population are likely to be in binary systems, and therefore out of reach to gamma-ray 
pulsation searches, there remains the potential for uncovering rare bright isolated MSPs 
in this region. 

Motivated by the arguments above, we used the donated computing cycles from
around 30,000 computing systems participating in the \textit{Einstein@Home}
distributed volunteer computing system \citep{Knispel2010, Allen2013+EatH, Pletsch2013+EAH} to
search for new gamma-ray pulsars in unidentified sources cataloged by
\citet[][hereafter \citetalias{2FIG}]{2FIG} and \citet[][hereafter
  \citetalias{4FGL}]{4FGL} in the inner $40^{\circ} \times
40^{\circ}$ around the GC. 
These searches resulted in the discovery of four new
pulsars, one of which is an isolated MSP, and another of which has the smallest 
angular separation of any known gamma-ray pulsar to the GC ($0.93^{\circ}$). Similar 
\textit{Einstein@Home} searches of pulsar-like \citetalias{4FGL} sources 
elsewhere on the sky have resulted in a further 10 discoveries\footnote{\url{https://einsteinathome.org/gammaraypulsar/FGRP1\_discoveries.html}}, which will be 
reported in a separate paper. 

This paper follows on from our previous descriptions of the
\textit{Einstein@Home} gamma-ray pulsar surveys in \citet[hereafter
  \citetalias{PaperI}]{PaperI} and \citet{PaperII}. It is organized as follows: in
Section~\ref{s:methods} we describe our preparation of the \textit{Fermi}-LAT
data, the sources that we targeted, and our gamma-ray pulsation searching
method; Section~\ref{s:results} presents the results of our pulsar searches and
follow-up investigations of the newly discovered pulsars; we discuss the
implications of our discoveries in Section~\ref{s:discussion}, with a summary
and conclusions in Section~\ref{s:conc}.

\section{Methods}
\label{s:methods}
\subsection{Gamma-ray source selection and data preparation}
\label{s:sources}
The input for the \textit{Einstein@Home} pulsation searches are lists of the
arrival times of gamma-ray photons detected by the \textit{Fermi-LAT}, with
estimated arrival directions from within a small (few degrees) patch on the sky. Unlike
searches for gamma-ray pulsations from known pulsars, which can detect
pulsations from very faint, even sub-threshold sources
\citep{Smith2019+ThousandPSRs}, our semi-coherent searches are only sensitive to
pulsars that are bright enough to also be detectable as individual point sources 
in \textit{Fermi}-LAT source catalogs through their
time-integrated emission (see Section~\ref{s:discussion}). Rather than searching over the whole sky (as is done
in \textit{Einstein@Home}'s searches for continuous gravitational waves from
unknown pulsars, e.g. \citealt{Steltner2023}), we therefore target our searches
at unidentified sources that have been detected through ``catalog'' analyses
that search for point sources of emission, e.g. \citetalias{4FGL} or \citetalias{2FIG}. 

As well as identifying gamma-ray point sources that may be pulsar candidates,
these catalogs provide a model describing the spectra of all detected sources,
from which we can derive photon probability weights
\citep{Bickel2008,Kerr2011,Bruel2019} that quantify the likelihood of each
photon being emitted by a targeted source, as opposed to by the diffuse
background, or a nearby unrelated point source. We use these to weight the
contribution of each photon to our pulsation detection statistics, to suppress
background photons without performing hard ``cuts'' that would be detrimental to
sensitivity.

To obtain a list of candidate sources for our searches, we performed an updated version of
the \citetalias{2FIG} analysis, using the same procedure but with a longer data set described below, and identified sources from \citetalias{4FGL} that
were deemed to be likely pulsar candidates by a Random Forest classification,
similar to that performed by \citet{SazParkinson2016}.

For our searches of \citetalias{2FIG} sources, we used the full duration of \textit{Fermi}-LAT data 
that was available at the time, approximately 8.7
years, including Pass 8 P8R2 data \citep{Pass8} between MJD 54682 (2008
August 04) and MJD 57847 (2017 April 04). This time span includes a 1-year 
interval\footnote{\url{https://fermi.gsfc.nasa.gov/ssc/observations/types/exposure/}} in which \textit{Fermi}'s observing strategy was modified to provide enhanced 
exposure towards the Galactic Center \citep{GCPointingMode}. We used \texttt{SOURCE}-class photons
and the \texttt{P8R2\_SOURCE\_V6} instrument
response functions (IRFs), with zenith angles smaller than
$90^{\circ}$. Photons from the inner-most $40^{\circ}\times40^{\circ}$ around the GC were
included, split into overlapping $8^{\circ}$-radius circular regions in which source
identification and spectral analyses were performed.

As described in \citetalias{2FIG}, these analyses were performed with two
different diffuse interstellar emission models: the \texttt{gll\_iem\_v06.fits}
model, used to construct the \citetalias{4FGL} catalogue (hereafter refered to
as the ``official'' (Off.)  model); and an ``alternate'' (Alt.) model based on
that of \citet{FermiCollab+GCExcess}. We also performed analyses with two
different lower energy bounds, 100\,MeV and 300\,MeV.

From the results of these analyses, sources were selected as being potential
pulsar candidates based on them having spectral properties consistent with the observed gamma-ray pulsar population. With the spectra
modeled as exponentially-cutoff power-law ($dN/dE \propto E^{-\Gamma}
\exp\left(-E/E_{\rm cut}\right)$), a source was deemed to be pulsar-like if it
had spectral index $0 < \Gamma < 2$ and cutoff energy $E_{\rm cut} < 10$~GeV.
From this initial list of possible pulsar candidate sources, we selected
sources from the $>100$~MeV and $>300$~MeV analyses that fulfilled the
pulsar-candidate criteria when fit using both the Off. and Alt. models, and had
similar spectra between the two analyses, as well as sources that were only
selected as pulsar candidates in the $>100$~MeV analysis with the Off. model. 

For each source, we used \texttt{gtsrcprob} to compute photon probability weights
for photons from within $5^{\circ}$ of the estimated source position, according to
the models produced by these analyses. We always used the spectral model
produced by the fits using the Off. model when computing photon probability
weights, unless the source was originally only identified as a pulsar candidate
with the Alt. model in the original \citetalias{2FIG} analysis of \citet{2FIG}. We included
as many of the highest-weighted photons as were necessary to reach 99\% of the
expected signal-to-noise ratio (S/N) from a source, up
to a maximum of 50,000 photons per source. This limit was imposed to avoid including a 
large number of low-weight photons that increase the computational cost of the search, 
which scales with the square of the number of photons, 
without contributing much to the recoverable S/N, 
which scales with the sum of the squared photon weights (see Equation~\ref{e:SN}). 
The 99\% expected S/N threshold was reached with fewer than 50,000 photons for half of 
the sources in our list, and the included photons contained at least 
90\% (80\%) of the expected total S/N for all but seven (three) sources. 

For sources from \citetalias{4FGL}, we used Pass 8 P8R3 data \citep{Pass8,Bruel2018+P305} up to MJD 58833 (2019 December
16), selecting \texttt{SOURCE} class photons with energies above $100$~MeV and using the \texttt{P8R3\_SOURCE\_V2} IRFs. We first included photons from within a
$20^{\circ}\times 20^{\circ}$ region around each source, and used
\texttt{fermipy} \citep{fermipy} to refit the spectra of the target source and all 4FGL sources
within $5^{\circ}$, as well as the normalisation and spectral indices of the
diffuse (\texttt{gll\_iem\_v07.fits}) and isotropic
(\texttt{iso\_P8R3\_SOURCE\_V2\_v1.txt}) background models. The resulting
spectral models were then used to compute photon weights using
\texttt{gtsrcprob} for photons within $5^{\circ}$ of the target source.

Finally, we removed sources that had too few weighted photons for us to
reasonably expect to detect pulsations with \textit{Einstein@Home} 
(determined from the expected pulsation signal-to-noise ratio, see Section~\ref{s:search}, Equation~(\ref{e:SN})). 

These selection criteria resulted in a list of 55 sources to search for pulsations, which are 
listed in Appendix~\ref{a:sources}, Table~\ref{t:sources}. Some of these were 
searched using two or three different sets of photon weights resulting from the different 
spectral analyses described above (i.e. using the \citetalias{2FIG} $>100$\,MeV, 
\citetalias{2FIG} $>300$\,MeV or \citetalias{4FGL} sky models) to attempt to mitigate the risk of
missing a new pulsar due to using photon weights computed from an inaccurate sky model.

For sources in which we detected significant gamma-ray pulsations, we produced
extended \textit{Fermi}-LAT data sets, covering data up to MJD 60595 (2024
October 12). These were prepared using the \texttt{P8R3\_SOURCE\_V3} IRFs \citep{Bruel2018+P305}, using the energy-dependent angular
and PSF (point spread function) event-type cuts described in
\citetalias{4FGL}. We computed photon weights using the spectral and spatial
model from \citet[][hereafter \citetalias{4FGL-DR4}]{4FGL-DR4}, with the
rescaled \texttt{gll\_iem\_v07} interstellar emission model associated to the 4FGL-DR4 catalog, and
\texttt{iso\_P8R3\_SOURCE\_V3\_v1.txt} isotropic background model.

\subsection{Gamma-ray pulsation searches}
\label{s:search}
We searched these sources for gamma-ray pulsations using the methods described 
in \citetalias{PaperI}. These methods are similar to the
``time-differencing'' technique developed by \citet{Atwood2006}, but incorporate
additional techniques developed to search for continuous gravitational waves
from spinning neutron stars in LIGO data
\citep{Balasubramanian1996+Metric,Owen1996+Metric,PletschAllen2009,Pletsch2011+slwin} and radio
pulsars \citep{Ransom2002+fourier}. Detailed descriptions of these methods can
be found in \citet{Pletsch2014+Methods} and \citet{Nieder2020+Methods}, which we
briefly summarize here.

Describing the rotational history of an isolated gamma-ray pulsar in the
\textit{Fermi}-LAT data requires an ephemeris ($\vec{\lambda}$) containing a
minimum of four parameters: the spin frequency ($f$), the spin-frequency
derivative ($\dot{f}$), and two positional parameters, which we take here to be
the equatorial Right Ascension ($\alpha$) and Declination ($\delta$).

In our searches, we use a simple constant spin-down phase model,
\begin{equation}
\begin{aligned}
  \phi\left(t;\vec{\lambda}\right) = f& \left(t_{\rm SSB}\left(t, \alpha, \delta\right) - t_{\rm ref}\right) \\
+& \frac{1}{2} \dot{f} \left(t_{\rm SSB}\left(t, \alpha, \delta\right) - t_{\rm ref}\right)^2\,,
  \label{e:phase_model}
\end{aligned}
\end{equation}
where $t_{\rm ref}$ is a reference epoch chosen to be close to the center of the
data span, and $t_{\rm SSB}(t, \alpha, \delta)$ is the estimated time that a
photon detected by the LAT at time $t$, propagating from a source at position
($\alpha,\delta$), would have arrived at the Solar-System Barycenter (SSB, an
approximately ``fixed'' point at the center of the Solar System). The conversions
between $t$ and $t_{\rm SSB}$ are known as ``barycentering'' corrections (with absolute values less than $\sim (1~\textrm{AU}/c) \approx 500$\,s), and are
required here to account for the position-dependent Doppler shift due to
\textit{Fermi}'s orbit around the Earth, and the Earth's orbit through the Solar
System (the Roemer delay), as well as much smaller general relativistic corrections for the Einstein delay ($\lesssim 2\,$ms) and Solar Shapiro delay ($\lesssim 100\,\mu$s) \citep[e.g.][]{tempo2}.

Pulsation strength in a set of $N$ weighted photon arrival times, for a given
ephemeris $\vec{\lambda}$, is often quantified by the $H$-test \citep{deJager1989,Kerr2011},
\begin{equation}
H(\vec{\lambda}) = \max_{M < 20}\left[4 - 4M + \sum_{n=1}^M \mathcal{P}_n\left(\vec{\lambda}\right)\right]
\end{equation}
which is a maximised sum over the harmonic Fourier powers, $\mathcal{P}_n$, 
\begin{equation}
  \mathcal{P}_{n}\left(\vec{\lambda}\right) = \frac{2}{\sum_j w_j^2} \left|\sum_{j=1}^N w_j e^{-2 \pi i n \phi(t_j;\vec{\lambda})}\right|^2\,,
\end{equation}
where $w_j$ is the probability weight for the $j$-th photon, and $t_j$ is its LAT arrival time.

In our searches, none of the four ephemeris parameters are known precisely enough in
advance to detect pulsations, and must therefore be searched over. We fill this
4-dimensional parameter space with a grid of search locations, whose density
depends strongly on the length of the data (we need to be more precise in each
parameter to accurately track the pulsar's rotation over longer periods of
time). 

An optimal spacing for this grid can be computed from the ``parameter space metric''
\citep{Balasubramanian1996+Metric,Owen1996+Metric}, which predicts how much
signal power will be lost due to a grid point not lying exactly at the
true ephemeris parameters. This metric can then be used to construct a grid that 
minimises the average loss in S/N for a given computing budget, to maximise the 
probability of detecting a signal \citep{Allen2021}. For gamma-ray pulsation searches, this metric
is derived in \citet{Pletsch2014+Methods} and \citet{Nieder2020+Methods}.

However, a grid dense enough to search for unknown pulsars in typical
unidentified \textit{Fermi}-LAT sources with arcminute positional uncertainties
using the $H$-test, or even the Fourier power in a single harmonic, is
computationally infeasible. Instead, we trade sensitivity for computational
efficiency by employing a ``semi-coherent'' detection statistic,
\begin{equation}
S_T\left(\vec{\lambda}\right) = \frac{1}{\kappa_{S_T}}\sum_{j = 1}^{N} \sum_{\substack{k = 1 \\ k \neq j}}^{N} w_j w_k e^{-2\pi i \left[\phi\left(t_j ;\vec{\lambda}\right) - \phi\left(t_k;\vec{\lambda}\right)\right]} W_{T}(t_j - t_k)\,,
\end{equation}
where $\kappa_{S_T}$ is a normalizing term, and $W_T(\tau)$ is a ``lag window''

\begin{equation}
W_{T}(\tau) = \begin{cases}
      1 & \text{for}\ |\tau| < T/2\,\\
      0 & \text{otherwise}\,
    \end{cases}\,,
\end{equation}
that only accepts photon pairs that are separated by less than half the
``coherence time'' $T$. The effect of this lag window is that the S/N is 
reduced, but our search only requires the phase model to be accurate over short
time scales $T$, rather than over the entire data span $T_{\rm obs}$, meaning we
can use a much sparser grid of search locations. In the limit where $T = T_{\rm obs}$, the statistic becomes the coherent Fourier power, $S_{T_{\rm obs}}(\vec{\lambda}) = \left(\mathcal{P}_1(\vec{\lambda}) - 2\right)/2$. 

For these searches, we used a coherence time of $T = 2^{22}\,{\rm s} \approx
48.5\,{\rm d}$, twice as long as that used in \citetalias{PaperI}. Our search
grid had a maximum ``mismatch'' (fractional loss in S/N due to an offset in
parameter space) of $15\%$ in each dimension. We searched two regions of
parameter space: one covering the majority of the known ranges of spin
parameters for young pulsars, with $f < 88$~Hz and $-\dot{f} <
10^{-10}$\,Hz\,s$^{-1}$; and another for MSPs with $88~{\rm Hz} < f < 1512~{\rm
  Hz}$ and $-\dot{f} < 10^{-13}$~Hz~s$^{-1}$. These regions were conservatively
chosen to also be sensitive to the second harmonic of nearly all of the fastest
spinning pulsars of each category. We searched circular sky regions around each
source, with radii 50\% larger than the semi-major axis of the 95\% confidence
region from the spectral analyses.

This parameter space was split into many ``work units'', each of which could be
processed by a typical personal computer within a few hours, which were then
distributed to the \textit{Einstein@Home} volunteer computers. At the end of
each work unit, the ten candidates with the highest detection statistic values
from the semi-coherent search were automatically followed-up with refined grids and more sensitive
statistics: first a semi-coherent refinement with $T=2^{23}$s, followed by a
fully-coherent search using $\mathcal{P}_1$. The ten candidates from the initial
semi-coherent search, and the ten best candidates from the coherent follow-up
stages were reported back to the central \textit{Einstein@Home} servers. The
most significant candidates from each source were then further refined on the
ATLAS cluster\footnote{\url{https://www.aei.mpg.de/atlas}} \citep{aulbert2008}, first using $\mathcal{P}_{1}$ with a smaller mismatch, and then
using the $H$-test, summing up to $M \leq 5$ harmonics. At this stage we
additionally computed the $H$-test with $f \to f/2$ and $\dot{f} \to \dot{f}/2$,
in case the second harmonic had been misidentified as the fundamental spin
frequency, which can happen for pulsars with double-peaked pulse profiles, since
for these pulsars $\mathcal{P}_2 > \mathcal{P}_1$.

Before launching our searches, we used the photon weights to estimate in advance
if it would be possible to detect pulsations from each source,
to avoid wasting computing time on sources that were too faint. As we argued in
\citetalias{PaperI} (Appendix A) and \citet{Nieder2020+Methods}
(Section 2.5), the expected semi-coherent spectral S/N,
\begin{equation}
  \label{e:SN}
\theta_{S}^2 = p^2 \left|\gamma\right|^2 \sqrt{\sum_{j = 1}^{N} \sum_{\substack{k = 1 \\ k \neq j}}^{N} w^2_j w^2_k W_{T}(t_j - t_k)}
\end{equation}
where $\gamma$ is the Fourier coefficient of the strongest harmonic of the pulse
profile (usually corresponding to the fundamental spin period, but sometimes the
second harmonic for pulsars whose pulse profiles have two similar peaks separated by half a rotation) and $p$
is the fraction of photons from the target source that are pulsed. Given a set
of photon weights from a target source, and with an assumed pulse profile shape,
we can therefore estimate the pulsed fraction necessary for our search to detect
pulsations above a certain S/N threshold. Gamma-ray pulsars tend to have $p
\approx 1$, but this fraction can appear to be smaller or even larger than unity
in the presence of mis-modeled background sources that manifest as inaccurate
photon weights. We choose optimistic assumptions to make this selection
conservative: we took $\left|\gamma\right| = 1$, corresponding to the narrowest
possible pulse profile (a delta function), and included sources where $\theta_{S}^2
= 8$ (approximately the threshold required for a source to reach the more
sensitive coherent follow-up stages in our search) could be reached with $p \leq
1$. The estimated minimum detectable pulsed fraction for each source is listed
in Table~\ref{t:sources}.

\section{Results}
\label{s:results}
We identified four new gamma-ray pulsars in the candidate lists returned by
\textit{Einstein@Home} volunteer computers. In Section~\ref{s:psrs} we describe
each of these pulsars in turn, and present the results of follow-up analyses
that are described in the next few sections.

These pulsars were all clear outliers in our list of candidate
$H$-tests (all had $H > 180$, while the most significant candidate that we do not believe to be a real pulsar signal had $H = 131$). 
In previous searches we have also identified pulsars that
only appeared in the semi-coherent search stages, due to e.g. timing noise or
glitches that prevented their power accumulating as expected in the coherent
searches. We therefore manually folded and attempted to refine the parameters of
marginal outlying semi-coherent candidates, but did not identify any new pulsars
in this way here.

After we began our search, two new radio MSPs,
PSR~J1624$-$39\footnote{\url{https://www.astro.umd.edu/\~eferrara/pulsars/GalacticMSPs.txt}} and PSR~J1730$-$0359 \citep{Fang2025},
were discovered in sources in our target list, 4FGL~J1624.3$-$3952 and 4FGL~J1730.4$-$0359, respectively, 
while ten more have likely associations with active galactic nuclei, supernova
remnants or pulsar wind nebulae, or radio/X-ray sources of unknown type listed
in \citetalias{4FGL-DR4}. We note these associations in
Table~\ref{t:sources}. The other 40 sources remain unassociated with any known
gamma-ray emitting sources detected at other wavelengths.

While we provide an estimate for the minimum pulsed fraction that each source
would need to have to be detectable in our survey in Table~\ref{t:sources}, we
caution against interpreting these as upper limits on pulsations, as these
estimates come with several caveats. First, there are types of pulsars that
would not be detectable in our searches, despite being brighter than these
thresholds. Most notably, we cannot detect pulsars in binary systems, which
includes the majority of MSPs due to the recycling process through which they
form, as the orbital Doppler shift makes our assumed phase model
inapplicable. While our semi-coherent detection statistic provides some
robustness to timing noise and small glitches from young pulsars, these
phenomena can also greatly reduce our sensitivity to pulsations. Second, the
photon weights themselves may also be inaccurate due to source confusion or
unmodeled sources near the targeted sources, and this can bias our estimated thresholds. Finally, while we search a conservatively large
range of sky positions around each target source, confusion between sources in
the Galactic plane can cause the source position to be inaccurate, putting the
true pulsar position outside our search region
\citep[e.g.][]{Clark2015+J1906}. While our searches can positively identify some
sources as being gamma-ray pulsars, these caveats mean that we cannot
confidently determine that a source is \textit{not} a gamma-ray pulsar.

\subsection{Gamma-ray Pulsation Timing}
\label{s:timing}
We performed a timing analysis using the 16-year \textit{Fermi}-LAT data set for
all four newly discovered gamma-ray pulsars to precisely estimate their
rotational and astrometric parameters. These analyses followed the methods used
in \citetalias{PaperI}, in which the timing parameters are varied in a
Markov-Chain Monte-Carlo (MCMC) process, to maximise the pulsation likelihood
according to a template pulse profile \citep{Ray2011}. For this work, we
included the extension to these methods from \citet{Nieder2019+J0952}, in
which the parameters of the template pulse profile are also varied within the MCMC process, to
marginalise over this source of uncertainty. We modeled the pulse profiles as a weighted
sum of wrapped Gaussian peaks, with the number of components chosen to minimise the Bayesian Information Criterion \citep{Schwarz1978+BIC}. 

Two pulsars in this work, PSR~J1736$-$3422 and PSR~J1748$-$2815, had significant
timing noise, i.e. their rotational phase significantly deviates from the
constant linear spin-down model of Equation~\ref{e:phase_model}. Timing noise is
common to all known types of pulsars, and is typically well-described by a noise
process with a steep ``red'' (i.e., larger amplitude at lower frequencies)
power-law spectrum
\citep[e.g.][]{Coles2011,Kerr2015+Timing,Parthasarathy2019+TN}. Because timing noise generally increases with spin period and spin-down rate \citep{SC2010}, it is usually not detectable in gamma-ray timing data for MSPs \citep{GPTA} and older non-recycled pulsars. We model timing noise in these pulsars as a stochastic noise process on
top of a cubic spin-down model (i.e., adding the next term, $\ddot{f}(t -
t_{\rm ref})^3 / 6$ to Equation~\ref{e:phase_model}). We use a reduced-rank
Fourier basis for this noise model, and include the corresponding Fourier
coefficients as timing model parameters, with amplitudes constrained by applying
a prior following a noise model spectrum whose parameters (``hyperparameters'')
we also fit for. This method is similar
to the ``photon-by-photon'' timing method of \citet{GPTA}, although here we
perform the fitting using the Gibbs sampling method used in
\citet{Thongmeearkom2024+TRAPUMRBs}. 

We model these noise processes using a power-law model with a 
low-frequency cutoff, with power spectra
\begin{equation}
    S(f) = \frac{A^2}{12 \pi^2} \left(\frac{f_{\rm c}}{1~{\rm yr}^{-1}}\right)^{-\gamma} \left(1 + \left(\frac{f}{f_{\rm c}}\right)^2\right)^{-\gamma/2}~{\rm yr}^{3}\,,
\end{equation}
parameterized by the amplitude $A$ at a reference frequency of $1~{\rm yr}^{-1}$, a corner frequency $f_{\rm c}$ and a spectral index $\gamma$. We applied log-uniform priors on $A$ and $0.1/ T_{\rm obs} < f_c < 10 / T_{\rm obs}$, and a uniform prior on $0.5 < \gamma < 15$.

\begin{figure*}
  \centering
  \includegraphics[width=\textwidth]{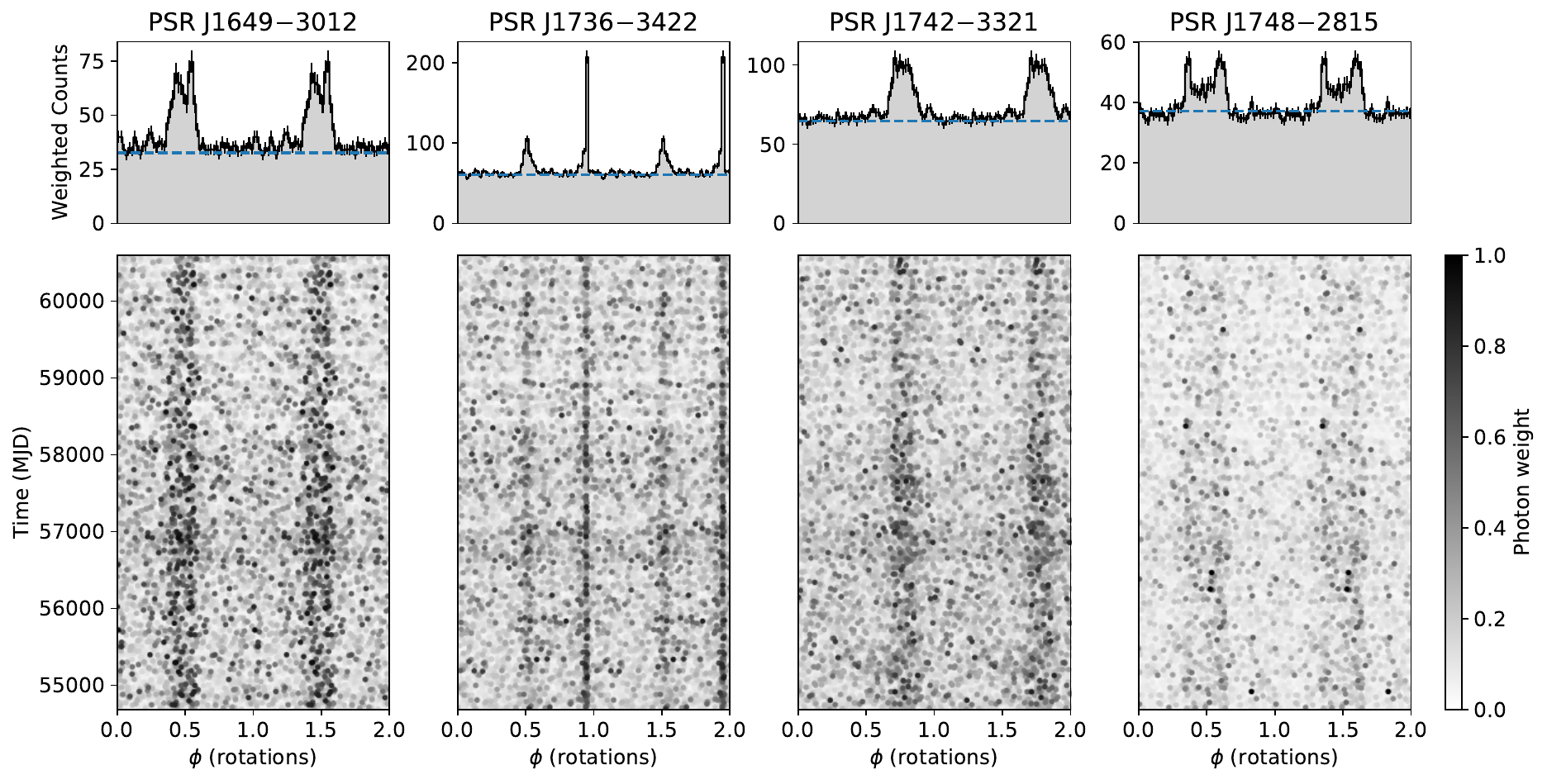}
  \caption{\label{f:phases} Photon phases and pulse profiles for the new
    gamma-ray pulsars discovered in this work. Lower panels show the phases for
    individual photons according to our best-fitting gamma-ray timing models,
    with the photon probability weights indicated by the grayscale. Upper panels
    show the integrated pulse profiles. The dashed blue lines show the estimated background level, calculated from the photon weights as
    $\sum_j w_j (1 - w_j) / N_{\rm bins}$ \citep{2PC}. Flux below these lines is attributed to the diffuse background or other nearby point sources. The period of enhanced exposure towards the Galactic Center is visible here as darker bands containing more photons between MJDs 56600 and 57000.
  }
\end{figure*}

Our gamma-ray timing solutions are given in Table~\ref{t:timing}, and we include
ephemerides compatible with the \texttt{tempo2} \citep{tempo2} and \texttt{PINT}
\citep{Luo2018+PINT} software as supplementary material. The resulting photon phases and
weighted pulse profiles are shown in Figure~\ref{f:phases}.

\begin{deluxetable*}{lcccc}
\tabletypesize{\footnotesize}
  \tablecaption{\label{t:timing} Gamma-ray timing solutions for newly discovered pulsars}
  \tablehead{
    \colhead{Parameter} &
    \colhead{PSR J1649$-$3012} &
    \colhead{PSR J1736$-$3422} &
    \colhead{PSR J1742$-$3321} &
    \colhead{PSR J1748$-$2815}
  }
  \startdata
  \hline
  \multicolumn{5}{c}{Timing parameters} \\
  \hline
  Data span (MJD) & \multicolumn{4}{c}{54681--60595} \\ 
  Reference epoch for spin frequency, $t_{\rm ref}$ & 56265.0 & 56757.0 & 56265.0 & 56265.0 \\
  Reference epoch for astrometry & 56265.0 & --- & --- & --- \\
  Solar system ephemeris & \multicolumn{4}{c}{DE405} \\
  Time scale & \multicolumn{4}{c}{TDB} \\
  \hline
  \multicolumn{5}{c}{Measured properties\tablenotemark{a}} \\
  \hline
  R.A. (J2000), $\alpha$ & $16^{\mathrm{h}}49^{\mathrm{m}}46\fs3164(8)$ & $17^{\mathrm{h}}36^{\mathrm{m}}10\fs90(2)$ & $17^{\mathrm{h}}42^{\mathrm{m}}11\fs58(3)$ & $17^{\mathrm{h}}48^{\mathrm{m}}31\fs02(1)$\\
  Decl. (J2000), $\delta$ & $-30^{\circ}12^{\prime}21\farcs21(6)$ & $-34^{\circ}22^{\prime}22(1)^{\prime\prime}$ & $-33^{\circ}21^{\prime}21(2)^{\prime\prime}$ & $-28^{\circ}15^{\prime}17(2)^{\prime\prime}$ \\
  Proper motion in R.A., $\mu_\alpha \cos\delta$ (mas yr$^{-1}$) & $-5(2)$ & --- & --- & ---  \\
  Proper motion in Decl., $\mu_\delta$ (mas yr$^{-1}$) & $7(11)$ & --- & --- & ---  \\
  Spin frequency, $f$ (Hz) & $292.01728830687(4)$ & $2.882409743(3)$ & $6.9780401641(1)$ & $9.9843315485(4)$\\
  Spin-frequency derivative, $\dot{f}$ (Hz s$^{-1}$) & $-1.1264(3)\times10^{-15}$ & $-5.4428(2)\times10^{-13}$ & $-6.1822(3)\times10^{-14}$ & $-3.52083(8)\times10^{-13}$ \\
  Second frequency derivative $\ddot{f}$ (Hz s$^{-2}$) & --- & $6(3)\times10^{-25}$ & --- & $-2.4(6)\times10^{-25}$  \\
  Glitch epoch, $t_{\rm g}$ (MJD) & --- & --- & 57401(19) & --- \\
  Glitch frequency increment, $\Delta f_{\rm g}$ (Hz) & --- & --- & $1.48(5)\times10^{-8}$ & --- \\
  Glitch spin-down increment, $\Delta \dot{f}_{\rm g}$ (Hz) & --- & --- & $-2.0(4)\times10^{-17}$ & --- \\
  \hline
  \multicolumn{5}{c}{Timing noise parameters\tablenotemark{b}} \\
  \hline
  Amplitude, $\log_{10} A$ & --- & $-10.8^{+0.7}_{-1.2}$ & --- & $-10.2^{+0.7}_{-1.1}$ \\
  Corner frequency, $\log_{10}\left(\frac{f_c}{1~{\rm yr}^{-1}}\right)$  & --- & $<-0.8$ & --- & $<-0.2$ \\
  Spectral index $\gamma$ & --- & $>6.0$ & --- & $>2.8$ \\
  Amplitude of 2nd component, $\log_{10} A_2$ & --- & $-8.1^{+0.3}_{-0.2}$ & --- & --- \\
  Corner frequency of 2nd component $\log_{10}\left(\frac{f_c}{1~{\rm yr}^{-1}}\right)$ & --- & $0.3^{+0.1}_{-0.1}$ & --- & --- \\
  \hline
  \multicolumn{5}{c}{Derived quantities\tablenotemark{c}} \\
  \hline
  Galactic longitude, $l$ (${}^\circ$) & $-8.042$ & $-5.675$ & $-4.149$ & $0.913$\\
  Galactic latitude, $b$ (${}^\circ$) & $+9.207$ & $-1.177$ & $-1.691$ & $-0.190$ \\
  Spin period, $P$ (ms) & $3.4244547841603(5)$ & $346.9319386(3)$ & $143.306713130(2)$ & $100.15693040(4)$\\
  Spin-down rate, $\dot{P}$ & $1.3209(3)\times10^{-20}$ & $6.5511(3)\times10^{-14}$ & $1.26963(6)\times10^{-15}$ & $3.53190(8)\times10^{-15}$\\
  Characteristic age, $\tau$ (yr) & $4\times10^9$ & $8.4\times10^{4}$ & $1.8\times10^{6}$ & $4.5\times10^{5}$\\
  Spin-down luminosity, $\dot{E}$ (erg s$^{-1}$) & $1.3\times10^{34}$ & $6.2\times10^{34}$& $1.7\times10^{34}$ & $1.4\times10^{35}$\\
  Surface $B$-field strength, $B_{\rm S}$ (G) & $2\times10^{8}$ & $4.82\times10^{12}$ & $4\times10^{11}$ & $6\times10^{11}$\\
  Light-cylinder $B$-field strength, $B_{\rm LC}$ (G) & $5\times10^{4}$ & $1\times10^{3}$ & $1\times10^{3}$ & $6\times10^{3}$\\
  \enddata
\tablenotetext{a}{Free parameters from our gamma-ray timing model, with 1$\sigma$ uncertainties on the final digits quoted in brackets}
\tablenotetext{b}{Since these timing noise hyperparameters typically have non-Gaussian posterior distributions, we quote 95\% confidence intervals, or 95\% confidence limits when an upper or lower bound cannot be determined from the data.}
\tablenotetext{c}{Due to the uncertain distances to these pulsars, we do not correct derived parameters for the Shklovskii effect or for relative acceleration in the Galactic potential. These effects are negligible for the three young pulsars, but are large and uncertain for the MSP J1649$-$2013 (see Figure~\ref{f:J1649_dist}).} 
\end{deluxetable*}

\subsection{Radio follow-up searches}
Targeting unidentified \textit{Fermi}-LAT gamma-ray sources with radio
telescopes is an efficient survey scheme for discovering new MSPs
\citep{Ray2012+PSC}. As such, many sources in our target list have already
undergone sensitive searches for radio pulsations \citep[e.g.,][]{Barr2013,SanpaArsa2016,Bhattacharyya2021,Clark2023+TRAPUM,Kerr2025+4MSPs}. 
The ephemerides that we
obtain through gamma-ray timing allow us to fold radio data to search for
pulsations from new gamma-ray pulsars, including in archival observations where we can
now achieve higher sensitivity than in the original searches performed when the
spin periods were unknown. Nevertheless, follow-up searches for radio pulsations from gamma-ray 
discovered pulsars have had a low detection rate \citep{PaperII,3PC,Ding2025}, 
likely reflecting the fact that pulsars' gamma-ray beams are observable 
from a wide range of viewing angles than the
comparatively narrow radio cones \citep[e.g.][]{Kalapotharakos2023}. 

The inner Galaxy is also being extensively studied at radio wavelengths, notably
at high sensitivity and angular resolution with MeerKAT by MMGPS \citep{MMGPS},
the SARAO MeerKAT Galactic Plane Survey \citep[SMGPS,][]{SMGPS} and the MeerKAT
legacy survey of the Galactic Center \citep{Heywood2022+MKGC}. Each of these
surveys provides imaging data from which we can place sensitive upper limits on
the presence of a radio counterpart to our gamma-ray MSPs based on
time-integrated flux.

MMGPS also includes pulsar search observations in three different frequency
bands (UHF at 544--1088~MHz, $L$-band at 856---1712~MHz, and S1 at
1968--2843~MHz). Data volumes from MMGPS pulsar search observations are too high
for archival storage, so we are unable to fold previous $L$-band observations of
these sources. However, during the UHF and $S$-band observations, summarised in Table~\ref{t:radio}, we formed
dedicated coherent tied-array beams on the best-fitting gamma-ray timing
positions for these pulsars, using 40 of the 44 antennas in MeerKAT's 1~km core.

For all pulsar-mode observations, we first cleaned the data of radio-frequency
interference (RFI) using either \texttt{filtool} \citep{PulsarX} for Murriyang
and MeerKAT observations, or \texttt{rfifind} \citep{presto} for GBT
observations, and then folded and searched the cleaned data over DMs using
\texttt{dspsr} \citep{dspsr} and \texttt{pdmp} \citep{psrchive}
(Murriyang/MeerKAT) or \texttt{prepfold} (GBT), maintaining full frequency
resolution. We searched a range of DMs from $0$\,pc\,cm$^{-3}$ up to twice
the maximum DM predicted for that line-of-sight by the NE2001 \citep{NE2001} and
YMW16 \citep{YMW16} Galactic electron density models.

No radio counterparts were detected for any of the new pulsars. 
We summarize both the archival and new radio observations that have been
performed towards our newly detected gamma-ray pulsars in
Table~\ref{t:radio}. For pulsar-mode observations, we estimate upper limits on the flux density using the pulsar
radiometer equation \citep{Handbook}, assuming a detection threshold at S/N = 8,
and that 70\% of the bandwidth is usable after
accounting for RFI and receiver roll-off. We assumed fractional pulse duty cycles of $W = 0.015 \sqrt{f/(1~{\rm Hz})}$ \citep{Kramer1998}, up to a maximum of $W = 0.1$ for the MSP~J1649$-$3012, where the flux density upper limit scales like $\sqrt{W/(1-W)}$.  We take the sky temperature from the
408~MHz all-sky map of \citet{Haslam1982}, reprocessed by
\citet{Remazeilles2015}, and assume it has a power-law spectrum with spectral
index $-2.6$. We estimate the flux density upper limit by integrating a nominal
power-law pulsar spectrum with index $-1.8$ \citep{Karastergiou2024+TPA} and the
sky temperature spectrum over the observing bandwidth, and report the flux
density limit at both the central observing frequency and extrapolated to a nominal frequency of 1400~MHz.

\begin{figure*}
  \centering
  \includegraphics[width=0.9\textwidth]{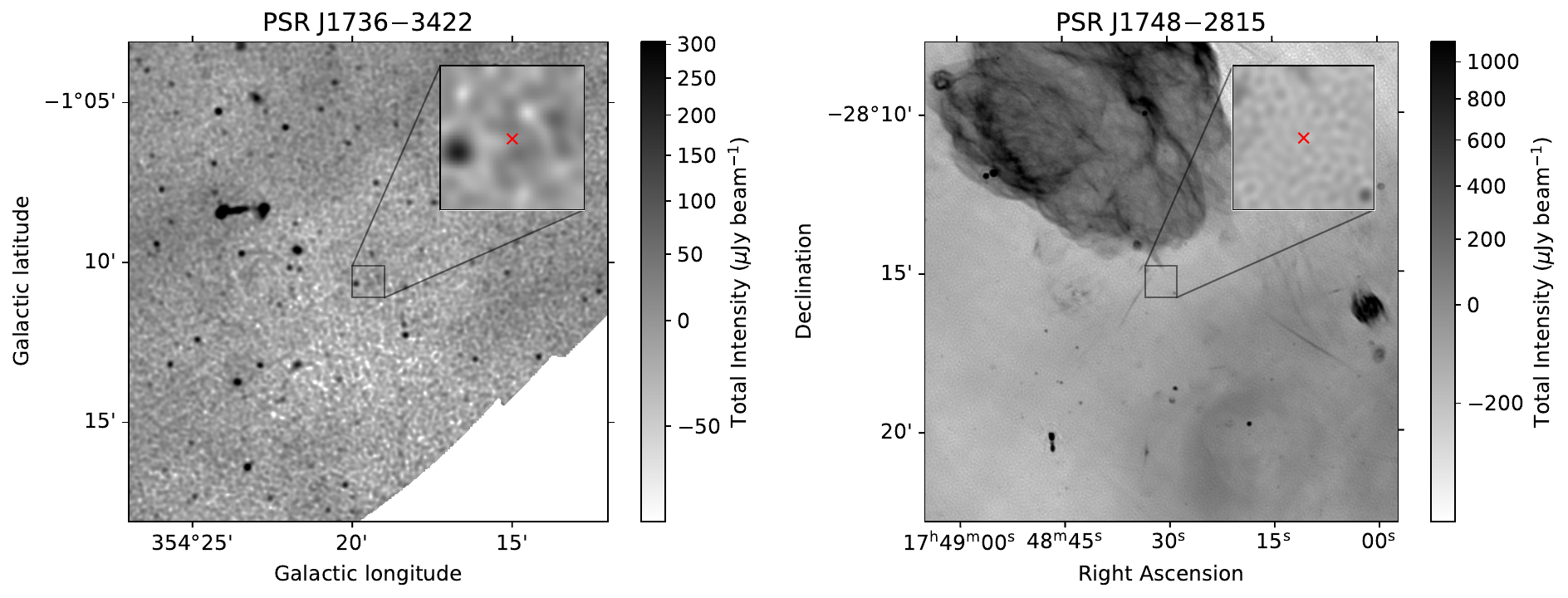}
  \caption{\label{f:radio_images} MeerKAT radio images at 1.28~GHz covering the locations of two
    new gamma-ray pulsars. No plausible radio counterparts are detected. Left panel: cutout from the SARAO MeerKAT Galactic
    Plane Survey \citep{SMGPS} covering the position of PSR~J1736$-$3422.  Right
    panel: cutout from the MeerKAT Galactic Center Mosaic
    \citep{Heywood2022+MKGC} covering the position of PSR~J1748$-$2815. Inset
    panels contain the $1\arcmin \times 1\arcmin$ region around the pulsars from
    which we estimated the RMS noise level to obtain flux-density upper limits,
    with the gamma-ray timing positions of the pulsars marked by red
    crosses. The positional uncertainties are smaller than these markers. The
    grayscale has the mean intensity subtracted, and a square-root
    normalisation applied, as indicated on the color bar.}
\end{figure*}

The positions of two new pulsars, PSRs~J1736$-$3422 and J1748$-$2815 were
covered by MeerKAT imaging surveys, shown in Figure~\ref{f:radio_images}. No
significant point sources consistent with the gamma-ray timing positions are
detected in these images. We estimated the RMS noise level from
$1\arcmin \times 1\arcmin$ regions around the gamma-ray timing positions of each
pulsar, after iteratively removing pixels that were above the $3\sigma$
level. We find $2\sigma$ continuum flux density upper limits of $39~\mu$Jy and
$53~\mu$Jy for PSRs~J1736$-$3422 and J1748$-$2815, respectively, similar to the most sensitive 1400~MHz pulsed flux density upper limits for these pulsars ($30\,\mu$Jy and $56\,\mu$Jy, respectively).

\begin{deluxetable*}{lcccccccc}
  \tablecaption{\label{t:radio}Radio pulsar search observations of newly-discovered pulsars}
  \tablehead{
    \colhead{Pulsar} & 
    \colhead{Telescope} &
    \colhead{Receiver/Backend} &
    \colhead{Epoch} &
    \colhead{Duration} &
    \colhead{Frequency} &
    \colhead{Max. DM} & 
    \colhead{$S_{\rm avg}$\tablenotemark{a}} & 
    \colhead{$S_{1400}$\tablenotemark{b}} \\
    & & & \colhead{(MJD)} & \colhead{(min)} & \colhead{(MHz)} & \colhead{(pc cm$^{-3}$)} & \colhead{($\mu$Jy)} & \colhead{($\mu$Jy)}
  }
  \startdata
  \hline
  J1649$-$3012 & GBT       & UHF/GUPPI  & 56725.5322 & 45 & $720$--$920$   & $400$ & $<53$ & $<20$ \\
               & Murriyang & H-OH/PDFB4 & 57697.2094 & 60 & $1241$--$1497$ & $400$ & $<87$ & $<84$ \\
               & GBT       & UHF/GUPPI  & 57781.5783 & 30 & $720$--$920$   & $400$ & $<65$ & $<25$ \\
               & MeerKAT   & UHF/FBFUSE & 60664.2180 & 8  & $544$--$1088$  & $400$ & $<91$ & $<35$ \\
  \hline
  J1736$-$3422 & MeerKAT   & S1/FBFUSE  & 60660.2534 & 20 & $1968$--$2843$ & $2300$ & $<11$ & $<30$ \\
                   &  MeerKAT & UHF/FBFUSE & 60664.2334 & 8  & $544$--$1088$ & $2300$ & $<85$ & $<32$ \\
  \hline
  J1742$-$3321 & MeerKAT   & UHF/FBFUSE  & 60664.2395 & 8 & $544$--$1088$ & $3000$ & $<100$ & $<38$ \\
  \hline
  J1748$-$2815 & Murriyang & UWL/Medusa & 59021.6701 & $3\times10.25$ & $704.5$--$4031.5$ & $3000$  & $<25$ & $<65$ \\
                   & Murriyang & UWL/Medusa & 59046.6033 & $3\times10.25$ & $704.5$--$4031.5$ & $3000$ & $<25$ & $<65$ \\
                   & MeerKAT   & S1/FBFUSE  & 60660.2370 & 20 & $1968$--$2843$ & $3000$ & $<21$ & $<56$ \\
                   & MeerKAT  & UHF/FBFUSE & 60664.2274 & 8 & $544$--$1088$ & $3000$ & $<307$ & $<116$ \\
  \enddata
  \tablenotetext{a}{Flux density upper limit at the central observing frequency, assuming a power-law pulsar spectrum with index $-1.8$, and a sky-temperature spectrum with index $-2.6$.}
  \tablenotetext{b}{Flux density upper limit, extrapolated to a reference frequency of 1400~MHz. }
\end{deluxetable*}

\subsection{Searches for Continuous Gravitational Waves}
The original motivation for the \textit{Einstein@Home} project was to carry out broad searches for continuous gravitational waves. Later, the \textit{Einstein@Home} searches were extended to search for new pulsars in radio and gamma-ray data. In turn, discoveries from these searches can be used for gravitational wave searches.

Gravitational wave emission is expected by non-axisymmetric neutron stars \citep[e.g.,][]{Brady1998} with the dominant signal frequency being equal to twice the rotational frequency of the star ($f_{\rm GW} = 2 f$). The ephemerides obtained from gamma-ray timing can be used to construct the gravitational wave search templates. If the gamma-ray timing solutions cover all epochs of the aLIGO observing runs, fully-coherent targeted searches can be executed, which yield maximum sensitivity.

We search for such signals using data from the first, second and third observing runs of the aLIGO detectors \citep{LIGOScientific:2019lzm, KAGRA:2023pio} from both PSRs J1649$-$3012\footnote{A preliminary search for this pulsar was presented in \citet{AshokThesis}. In this paper we present search results using an updated timing solution.} and J1748$-$2815 as both objects are expected to produce signals within the detector band. 

We use two distinct analysis procedures which yield similar results. The first is a frequentist approach, based on the multi-detector maximum-likelihood $\mathcal{F}$-statistic \citep{Cutler2005}. The second is a Bayesian $\mathcal{F}$-statistic-based approach detailed by \citet{Ashok2024}. Both methods coherently combine data from the Hanford and Livingston aLIGO detectors.

The results are consistent with a non-detection. At twice the rotational frequency, we set the $\mathcal{F}$-statistic 95\% confidence upper limits on the intrinsic gravitational wave amplitude $h_{0}$ and the corresponding Bayesian 95\% credible interval upper limits on the intrinsic amplitudes, summarized in Table~\ref{t:cw_results}.

Additionally, since the gravitational wave emission may be mismatched from twice the rotational frequency, we performed an $\mathcal{F}$-statistic search in a band $\Delta f_{\rm GW} = 4 \times 10^{-3} \times f_{\rm GW}$ around each pulsar and likewise for the spindown $\dot{f}$ consistent with previous searches \citep{Abbott2022, Ashok_2021}. The search was divided into multiple 10-mHz wide sub-bands and upper limits on $h_{0}$ are set in each band. We obtain a mean upper limit value of $h_{0}$ across all 10-mHz bands of $2.6 \times 10^{-26}$ and $6.2 \times 10^{-25}$ for J1649$-$3012 and J1748$-$2815 respectively, larger values than the targeted search upper limits due to the ``trials factor'' effect of searching over many waveform templates. We find no evidence of gravitational wave emission in these additional extended parameter space searches as the maximum $\mathcal{F}$-statistic value in each band is consistent with Gaussian noise.

Our upper limits on $h_{0}$ can be expressed as upper limits on the ellipticity $\epsilon$ of the pulsars \citep{Abbott_2019}. This is,
\begin{multline}
    \epsilon = 2.36 \times 10^{-6} \times \left(\frac{h_{0}}{10^{-25}}\right) \left(\frac{10^{45}~{\rm g~cm}^{2}}{I_{\rm zz}}\right) \\ \times \left(\frac{100~{\rm Hz}}{f_{\rm rot}}\right)^{2} \left(\frac{d}{1~{\rm kpc}}\right)\,,
\end{multline}
where $I_{\rm zz}$ is the moment of inertia about the spin axis, d is the distance to the pulsar and $f_{\rm rot}$ is the rotational frequency. The ellipticity upper limits can be calculated as a function of distance to the pulsar. The calculated ellipticity upper limit values as a function of distance, using the frequentist upper limits on the intrinsic gravitational wave amplitude and the canonical value of $I_{ \rm zz} = 10^{45}$g~cm$^{2}$ are shown in Table~\ref{t:cw_results}.

\begin{deluxetable*}{lcccccc}
  \tablecaption{\label{t:cw_results} Continuous Gravitational Wave Upper Limit Results}
  \tablehead{
    \colhead{Pulsar} & 
    \colhead{$f_{\rm GW}$} & 
    \colhead{$h_{0}^{95\%}$} &
    \colhead{$h_{0}^{95\%}$ (Bayes.)} &
    \colhead{$h_{0}^{\rm spdwn} \times \left(d / 1 {\rm kpc}\right)$} &
    \colhead{$\epsilon^{95\%} \times \left(1 {\rm kpc} / d\right)$} & 
    \colhead{$h_{0}^{95\%}/h_{0}^{\rm spdwn}$} \\
    & [Hz] & & & & & (at $d= 1 {\rm kpc}$)
  }
  \startdata
  \hline
  J1649$-$3012 & $\approx 584.03$ & $5.5 \times 10^{-27}$ & $4.2 \times 10^{-27}$ & $1.6 \times 10^{-27}$ & $1.5 \times 10^{-8}$ & $3.4$ \\
  J1748$-$2815 & $\approx 19.97$ & $1.2 \times 10^{-25}$ & $1.0 \times 10^{-25}$ & $1.5 \times 10^{-25}$ & $2.8 \times 10^{-4}$ & $0.8$\\
  \enddata
\end{deluxetable*}

It was possible albeit unlikely that our searches would have detected a gravitational wave. How likely these searches would have yielded a detection and the physical significance of our results can be assessed by taking the ratio between our measured gravitational wave amplitude upper limits and the theoretical maximum ``spindown limit" gravitational wave amplitude $h_{0}^{ \rm spdwn}$; that is the gravitational wave amplitude produced by the pulsar assuming that all of the rotational energy was channeled into gravitational wave emission. If $h_{0}^{95\%} / h_{0}^{\rm spdwn}$ is less than 1, then the upper limits are informative. Assuming the most optimistic case in which each pulsar is located at the minimum possible distance of 1 kpc from Earth (see next subsection), we compute $h_{0}^{\rm spdwn}$. The corresponding results using the frequentist upper limits are shown in Table~\ref{t:cw_results}.

\subsection{Estimating gamma-ray pulsar distances}
Distances for new pulsars are typically estimated by comparing the radio
dispersion measure to the line-of-sight electron column density predicted by the
NE2001 \citep{NE2001} and YMW16 \citep{YMW16} models. However, most pulsars
discovered in gamma-ray searches appear to be ``radio quiet'', and this includes
all four of the new pulsars discovered here.

In the absence of a radio detection and resulting dispersion measure, we can
only crudely estimate the distance to the new pulsars by using the measured
gamma-ray fluxes and appealing to the estimated spin-down power budget. In
\citetalias{3PC}, the gamma-ray efficiencies $\eta = L_{\gamma} / \dot{E}$
implied by the distances ($d$), gamma-ray luminosities ($L_{\gamma}$) and
spin-down powers ($\dot{E} = 4 \pi^2 I_{\rm zz} f \dot{f}$, for an assumed
moment of inertia $I_{\rm zz} = 10^{45}$g~cm$^{2}$) of the population of
gamma-ray MSPs were found to lie within the range $0.01 \lesssim \eta \lesssim
1$. \citet{Kalapotharakos2023} found that pulsars' gamma-ray beams sweep out a
non-isotropic pattern on the sky in which most viewing angles receive more flux than they would if emission was isotropic, meaning that efficiencies calculated under the assumption of
isotropic emission are usually over-estimates. This also means that apparent
efficiencies $\eta > 1$ are possible, but no pulsar in \citetalias{3PC} was
found to significantly exceed this level, so we use this as an approximate upper
limit on possible luminosities. Motivated by gamma-ray emission models that
predict $L_{\gamma} \propto \dot{E}^{1/2}$ \citep{Arons1996}, a ``heuristic'' efficiency, $\eta_h
= (\dot{E}/10^{33}\,\mathrm{erg\,s})^{-1/2}$ has been found to provide a
reasonable prediction for the majority of the gamma-ray pulsar population \citepalias{3PC}. For the four new pulsars found here, these heuristic efficiencies range from 10\% to 30\%. We
therefore use the observed gamma-ray fluxes from \citetalias{4FGL-DR4} and these
typical efficiencies to estimate likely and maximum distances to the new pulsars.

\subsection{New Pulsars}
\label{s:psrs}
In the following sections we give individual details and results from our
follow-up investigations of the newly-discovered gamma-ray pulsars.

\subsubsection{PSR J1649$-$3012}
With a spin period of $3.4$~ms, this is the third isolated MSP to be found by
\textit{Einstein@Home} \citep{Clark2018+EAHMSPs}. We were motivated to search
inner-Galaxy sources to try to identify MSPs from the putative bulge population,
and so we wish to investigate if PSR~J1649$-$3012 belongs to this
population. This requires us to estimate the distance to the pulsar, to
determine if it might reside in the Galactic bulge, at a distance of
6--10~kpc from the Solar System, or from the local Galactic disk population closer to Earth.

The gamma-ray source in which we discovered this pulsar was long considered
a strong MSP candidate, and as such has been targeted by previous radio surveys\footnote{A gamma-ray source, J1649$-$3004, near this position was also targeted with the Giant Metrewave Radio Telescope (GMRT) by \citet{Bhattacharyya2021}, however the timing position of the new pulsar is outside the primary beam of this observation.}
of unidentified \textit{Fermi}-LAT sources with the Green Bank Telescope (GBT)
\citep{SanpaArsa2016}, and Murriyang \citep{Kerr2025+4MSPs}. We folded
these archival observations, as well as a dedicated coherent beam formed on our best-fitting timing position during the MMGPS UHF observation, but
did not detect any pulsations. This makes PSR~J1649$-$3012 only the third known
``radio-quiet'' gamma-ray MSP \citep{Clark2018+EAHMSPs,Nieder2020+J1653} 
that has been subjected to sensitive radio searches without detection. 
The expected dispersion smearing and scattering timescales for the most sensitive GBT observations
are smaller than 10\% of the rotation period for all DMs up to the maximum expected DM (248\,pc\,cm$^{-3}$ from NE2001) for this line-of-sight, and so this non-detection is unlikely to be due to dispersive smearing or scattering. 
The lack of radio detection could be explained by either the pulsar having a 
low intrinsic radio luminosity, or a radio beam that misses our line-of-sight. 

\begin{figure}
  \centering
  \includegraphics[width=0.95\columnwidth]{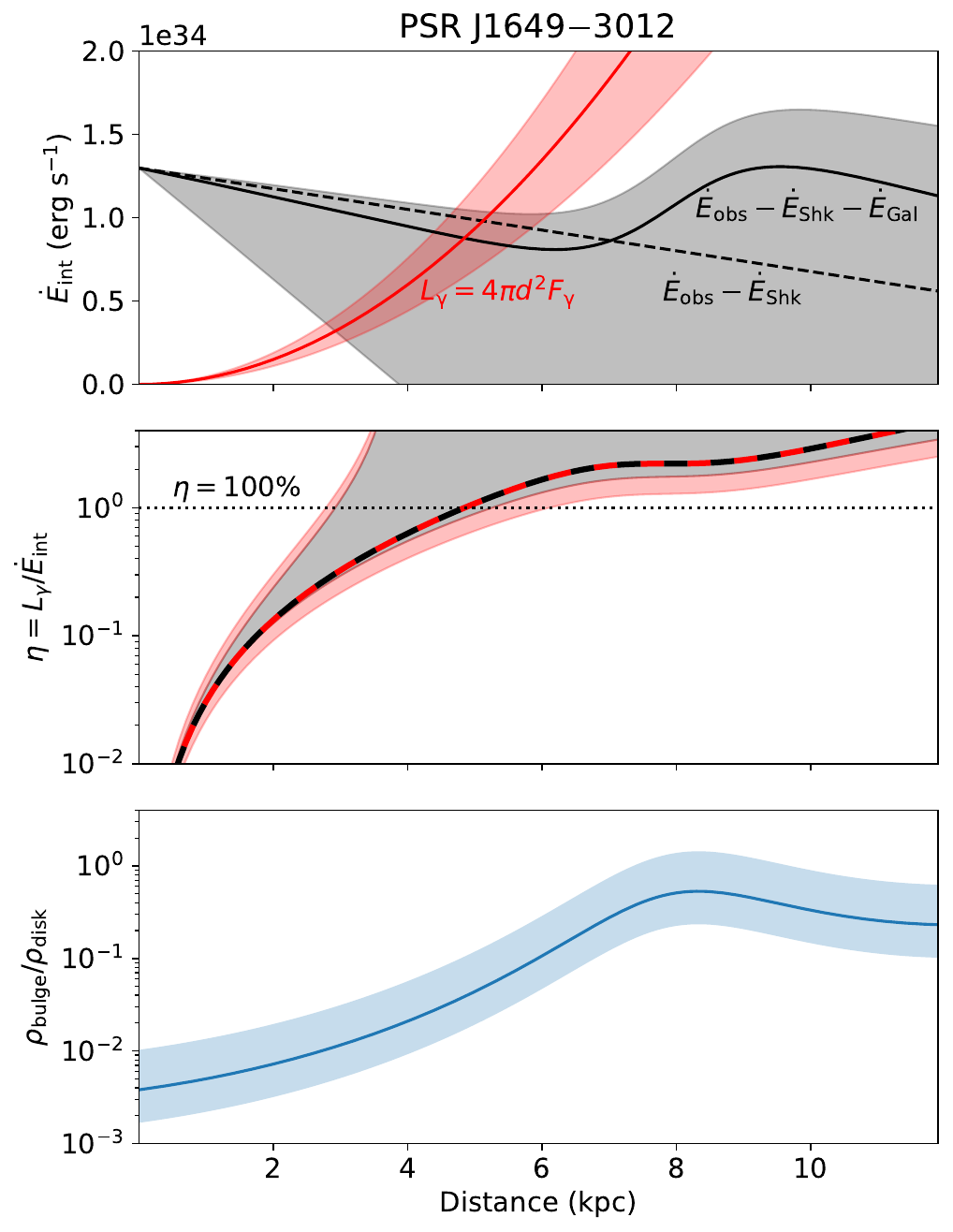}
  \caption{
    \label{f:J1649_dist}
    Distance dependence of inferred properties of PSR~J1649$-$3012. The top
    panel shows the intrinsic spin-down luminosity ($\dot{E}_{\rm int}$), after correcting for the Shklovskii effect according to our best-fitting timing model (dashed black line) and also the acceleration in the Galactic potential (solid black line). The grey shaded region shows 1-$\sigma$ uncertainties on the intrinsic spin-down luminosity, propagated from the uncertainty on the total proper motion, which we assume to be much larger than the uncertainty in the Galactic acceleration. The gamma-ray
    luminosity ($L_{\gamma}$) and 1-$\sigma$ uncertainties according to the observed energy flux ($G_{\gamma}$)
    from \citetalias{4FGL-DR4} are shown by the red line and shaded
    region. The middle panel shows the corresponding gamma-ray efficiency, with the ratio of the red and black lines from the top panel shown by the dashed red and black line. The relative contributions to the efficiency uncertainty from the proper motion and observed flux uncertainties are shown by grey and red shaded regions, respectively. The
    dotted black line shows a loose upper limit corresponding to 100\%
    efficiency. The bottom
    panel shows the ratio between the predicted density of MSPs in the putative
    Galactic bulge population from the model of \citet{Calore2016} and in the
    Galactic disk according to the model of \citet{Levin2013+GalMSPs}.}
\end{figure}

The observed spin-down rate, $\dot{f}_{\rm obs}$ for this pulsar results in an
estimated spin-down luminosity, $\dot{E} \approx 1.3 \times 10^{34}~{\rm erg
  s}^{-1}$. However, we must correct the observed spin-down rate for relative
acceleration between the pulsar and the SSB. One source of apparent acceleration
is the so-called Shklovskii effect \citep{Shklovskii}, whereby initially tangential proper motion
picks up an increasingly radial component. The magnitude of this acceleration is
distance-dependent, $\Delta\dot{f}_{\rm Shk} = f \mu^2 d / c$, where $\mu =
\sqrt{\mu_{\delta}^2 + \mu_{\alpha}^2 \cos^2 \delta}$ is the total proper
angular speed. From our gamma-ray timing solution, we obtain a marginal ($2.5\sigma$)
detection of proper motion in the R.A. direction ($\mu_\alpha\cos\delta = -5 \pm
2\,{\rm mas}\,{\rm yr}^{-1}$), but motion in the Decl. direction is less precisely
estimated ($\mu_\delta = 7 \pm 11\,{\rm mas}\,{\rm yr}^{-1}$). As a result, the Shklovskii
contribution to the observed spin-down rate is rather uncertain: this effect could be almost negligible, 
or it could account for all of the observed spin-down at distances greater than around 4~kpc. 
Relative acceleration due to the Galactic potential, which we
evaluate using the \texttt{PJM17\_best.Tpot} model of \texttt{GalPot}
\citep{GalPot}, is a smaller effect at small distances, but at distances $>7\,{\rm kpc} $ 
its contribution to the observed spin-down rate becomes more significant and positive, 
and therefore allows for larger intrinsic spin-down luminosities. 
We show the distance-dependence of the inferred intrinsic spin-down luminosity 
and gamma-ray efficiency in Figure~\ref{f:J1649_dist}. We
find that the $\eta \approx 1$ level is possible for all distances $d \gtrsim
2.5$\,kpc, even including $d \gtrsim 8$\,kpc if the total proper motion is on the
lower end of our constraints. However, the range of MSP efficiencies seen in \citetalias{3PC} of $0.01$--$1$ suggests a distance of $1\,\mathrm{kpc}\lesssim d \lesssim
6\,\mathrm{kpc}$ is more likely here.

\subsubsection{PSR J1736$-$3422}
With a characteristic age of 84~kyr, this is the youngest pulsar discovered in
this work. Due to the double-peaked shape of its pulse profile,
\textit{Einstein@Home} only detected the second harmonic as a significant
candidate, but our automatic folding at half the original spin frequency
revealed significant excess Fourier power in odd-numbered harmonics, clarifying the
fundamental spin frequency as $f \approx 2.88$~Hz.

\begin{figure*}
  \centering
  \includegraphics[width=0.9\textwidth]{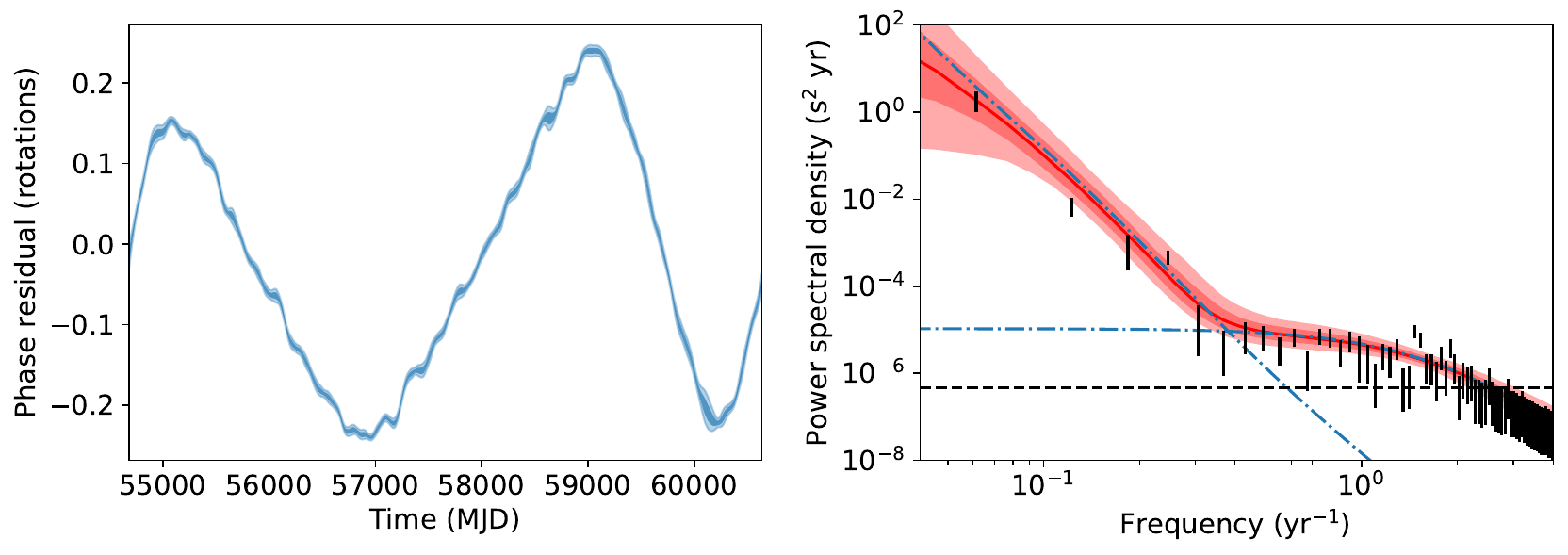}
  \caption{
    \label{f:J1736_TN}
    Illustration of timing noise in PSR~J1736$-$3422. The left panel shows the estimated variations in the rotational phase, after subtracting a cubic spin-down model, with the 2-$\sigma$ confidence interval indicated by the blue shaded region. The right panel shows the resulting power spectrum of these phase variations. The black error bars show the $1\sigma$ posterior uncertainties on the Fourier powers, marginalised over the underlying
    timing model and template pulse profile shape. The blue dot-dashed lines show the best
    fitting models for the two noise components, while the red line and shaded
    region shows the best-fitting model and uncertainty regions for the full noise
    model prior. The estimated white-noise level is shown by the black dashed line. The posterior distributions for high frequencies are all below this level, and closely follow the priors as a result.  }
\end{figure*}

This pulsar exhibits significant timing noise, illustrated in Figure~\ref{f:J1736_TN}, that contains detectable features even at relatively high frequencies (of a few per year). We were unable to
obtain an acceptable model to the observed timing noise power spectrum (shown in
Figure~\ref{f:J1736_TN}) using a single cut-off power-law spectrum. This either
returned a steep spectrum (with spectral index of $\gamma > 6$) that left
excess power at intermediate frequencies of $0.5$--$2$~yr$^{-1}$ (and visible
residual trends in the resulting photon phases), or a shallow spectrum (with
$\gamma \approx 3.5$) that accounted for both high- and low-frequency power, but
over-estimated the power at all intermediate frequencies. We therefore added a
second smoothly broken power-law component to our model. This two-component
model resulted in a more typical steep component at low frequencies (with
$\gamma > 6$ at 95\% confidence), and a low-amplitude flat component with
corner frequency $f_c = 2.0 \pm 0.6$~yr$^{-1}$. The corner frequency of the steep component is
constrained only to be similar or below one cycle in the 16~yr observation time,
meaning that this component is well modeled by a pure power-law. The spectral index of
the high-frequency component is not constrained by the data, since this
component falls quickly below the white noise level at frequencies above the
corner frequency, so we fixed this parameter at $\gamma_2 = 6$.

This two-component timing noise spectrum is, to our knowledge, highly unusual
for a gamma-ray pulsar. Correlated phase and 
pulse-profile variations have been detected in many radio pulsars \citep{Lyne2010,Brook2016,Lower2025},
including in the gamma-ray pulsars J0742$-$2822 and J1830$-$1059 that have similar spin-down luminosities to PSR~J1736$-$3422, which 
\citet{Lyne2010} found could be explained by quasi-periodic switching between 
two different magnetospheric states, each with a distinct spin-down rate and radio pulse 
profile shape. 
By simulating timing noise processes according to that model, 
we found that we could produce power spectra that were qualitatively similar to the spectrum
we measure from J1736$-$3422, if the switching timescale is around 100~d and varies by around 25\%, 
and the spin-down rate differs by around $2\times10^{-15}$~Hz~s$^{-1}$ between the high- and low- states. These values are comparable to those in the \citet{Lyne2010} sample,
making this a plausible explanation for the observed high-frequency timing noise structure. 

However, this hypothesis is difficult to test further. In the absence of detectable radio pulsations, we cannot test for correlated
changes in the radio pulse profile, of the kind seen in \citet{Lyne2010}. The possible switching timescale 
is also too short for us to sensitively check for similar shape variations in the gamma-ray pulse profile. 
Despite several of the magnetospheric-switching radio pulsars also being detected in gamma rays, 
flux and pulse-profile variability has only been detected from one gamma-ray pulsar, PSR~2021$+$4026
\citep{Allafort2013+J2021,Wang2024+J2021}. That pulsar does appear to transition between long-lived
gamma-ray flux/spin-down states, but the spin-down variations are orders 
of magnitude larger than we observe here, suggesting 
much larger global changes in the magnetosphere are at play there \citep{Fiori2024+J2021}.

PSR~J1736$-$3422 is one of only two young ($<100$~kyr) pulsars within one degree
of the supernova remnant (SNR) G354.8$-$0.8, with \citet{Espinoza2022+J1734}
concluding that the younger and more nearby PSR~J1734$-$3333 ($21\arcmin$ from the SNR) 
cannot be
associated with this remnant. The age and distance to this SNR are unclear, but
we can use the characteristic age and estimated distance for PSR~J1736$-$3422 to
investigate a possible association. The angular separation of $40\arcmin$
between this pulsar and the SNR would require a proper motion of around
30\,mas\,yr$^{-1}$, which would not be detectable to our gamma-ray timing
analysis\footnote{The detection of the much smaller proper motion from the 
MSP J1649$-$3012 was possible because the uncertainty on astrometric parameters
scales inversely with the pulsar's spin frequency.}. From the gamma-ray flux reported in \citetalias{4FGL-DR4}, and using
the heuristic efficiency, we find that the distance to PSR~J1736$-$3422 is
likely to be between 1 and 4 kpc. The required proper motion therefore
corresponds to a plausible kick velocity of $(d / 1\,{\rm
  kpc})\times140$\,km\,s$^{-1}$. There is no sign of any radio emission in the
SMGPS image covering this pulsar (see Figure~\ref{f:radio_images}), either as a point source or from a bow-shock that would provide evidence that the pulsar is traveling with high velocity
through the interstellar medium. The association between this pulsar and the SNR 
therefore remains plausible, but speculative. 

\subsubsection{PSR J1742$-$3321}
This is a middle-aged (2~Myr) pulsar found in a \textit{Fermi}-LAT source lying
4.5$^{\circ}$ from the GC. Based on the estimated spin-down power of
$1.7\times10^{34}~$erg~s$^{-1}$, we estimate the distance to lie in the range of
$0.8~{\rm kpc} < d < 3~{\rm kpc}$. 

This pulsar also exhibits deviations from a linear spin-down model, but in this
case we were able to model this as a small glitch occuring in 2016 January, with
a fractional frequency increment $\Delta f_{\rm g} / f \approx 2 \times
10^{-9}$. This glitch size is very typical of the low-magnitude glitches in the
bi-modal glitch distribution found by \citet{Basu2022+Glitches}. While this
glitch happened within the span of our search data set, it had very little
impact on our search sensitivity; the magnitude is small enough that the pulsed
signal still remains well in phase over the coherence time of our semi-coherent
search (during which time a phase shift of only $\Delta \phi = \Delta f_{\rm g}
T = 0.062$~rotations would accumulate).

\subsubsection{PSR J1748$-$2815}
At an angular separation of $0.93^{\circ}$, this has the smallest known angular separation of any 
gamma-ray pulsar to the GC, and was found in the closest source to the GC in our target
list. The next closest gamma-ray pulsar, PSR~J1747$-$2958, ($1.09^{\circ}$ from the GC) 
was originally found
in a radio search of the ``Mouse'' nebula \citep{Camilo2002+Mouse}.

This location poses a particular challenge for gamma-ray spectral and spatial
modelling: this pulsar and the GC are separated by less than the 68\%
containment angle of the LAT point spread function\footnote{\url{https://www.slac.stanford.edu/exp/glast/groups/canda/lat_Performance.htm}} for all photon energies below 
800~MeV. Flux from this pulsar therefore represents a contaminating factor for
measuring the spectral and spatial properties of the GC excess at low energies,
and vice versa. Additionally, this position within the Galactic plane has an
unusually high contribution from the diffuse interstellar emission; this pulsar
lies adjacent to the two brightest pixels in the \texttt{gll\_iem\_v07.fits}
interstellar emission model used to construct the \citet{4FGL} catalog and its
incremental releases. As a result of this, the estimated significance and
spectrum of the gamma-ray flux from this pulsar depends strongly on the chosen
background model and analysis methods. Despite appearing as a highly significant
source in the earlier \textit{Fermi}-LAT Third Source Catalog \citep{3FGL} and in
the \citetalias{2FIG} catalog used here, this pulsar was not even identified as
a significant point source in \citetalias{4FGL-DR4}. This is likely due to a
combination of flux from this source being misattributed to the diffuse
background model or to other nearby sources, and the weighted log-likelihood
method used in \citetalias{4FGL} to downweight the significance of residual flux in regions
of the sky where systematic uncertainties in the background model dominate over
statistical uncertainties.

Our discovery of gamma-ray pulsations from this source therefore provides a
valuable ``anchor'' for gamma-ray modelling of the GC region, as we can use the
pulsed nature of the emission from this source to more confidently disentangle
this source from its surroundings. We do this in a multi-step process: first, we applied
the ``simple weighting'' scheme of \citet{Bruel2019} to compute approximate photon
weights without requiring an accurate gamma-ray sky model, using a reference energy
of $10^{4.2}~{\rm MeV} \approx 16~{\rm GeV}$ that maximized the weighted $H$-test. Next, we used these
initial weights to obtain a preliminary timing solution, which we could then use
to apply the ``model weighting'' scheme from \citet{Bruel2019}, in which the
parameters of an exponentially-cutoff power-law spectrum model for the pulsar
are varied to maximise the resulting weighted $H$-test. These updated weights
were then used to perform the full timing analysis, including a red timing noise
model, as described in Section~\ref{s:timing}. Finally, using this resulting
timing solution, we computed the phase of \texttt{SOURCE}-class photons, according to the \texttt{P8R3\_SOURCE\_V3} IRFs, detected 
up to MJD 60524 (2024 August 02), 
with energies $63~{\rm MeV} < E < 31.6~{\rm GeV}$, and with zenith angles $<105^{\circ}$. We separated these into on- and off-pulse phase ranges by employing
the Bayesian Blocks algorithm \citep{Scargle1998+BB} on the weighted pulse
profile, and binned these in energy and angular direction to obtain
energy-dependent on- and off-pulse counts maps. The on/off-pulse phase intervals determined 
by the Bayesian Blocks method depend on the prescription used for computing the 
uncertainty on the weighted photon counts in each phase bin, and is either $0.26$--$0.65$ 
(for Poisson uncertainties, $\sqrt{\sum_j w_j^2}$) or $0.34$--$0.65$ 
(for uncertainties of $\sqrt{1 + \sum_j w_j^2}$, as used in e.g. \citealt{2PC}). 

\begin{figure*}
\centering
\includegraphics[width=0.9\textwidth]{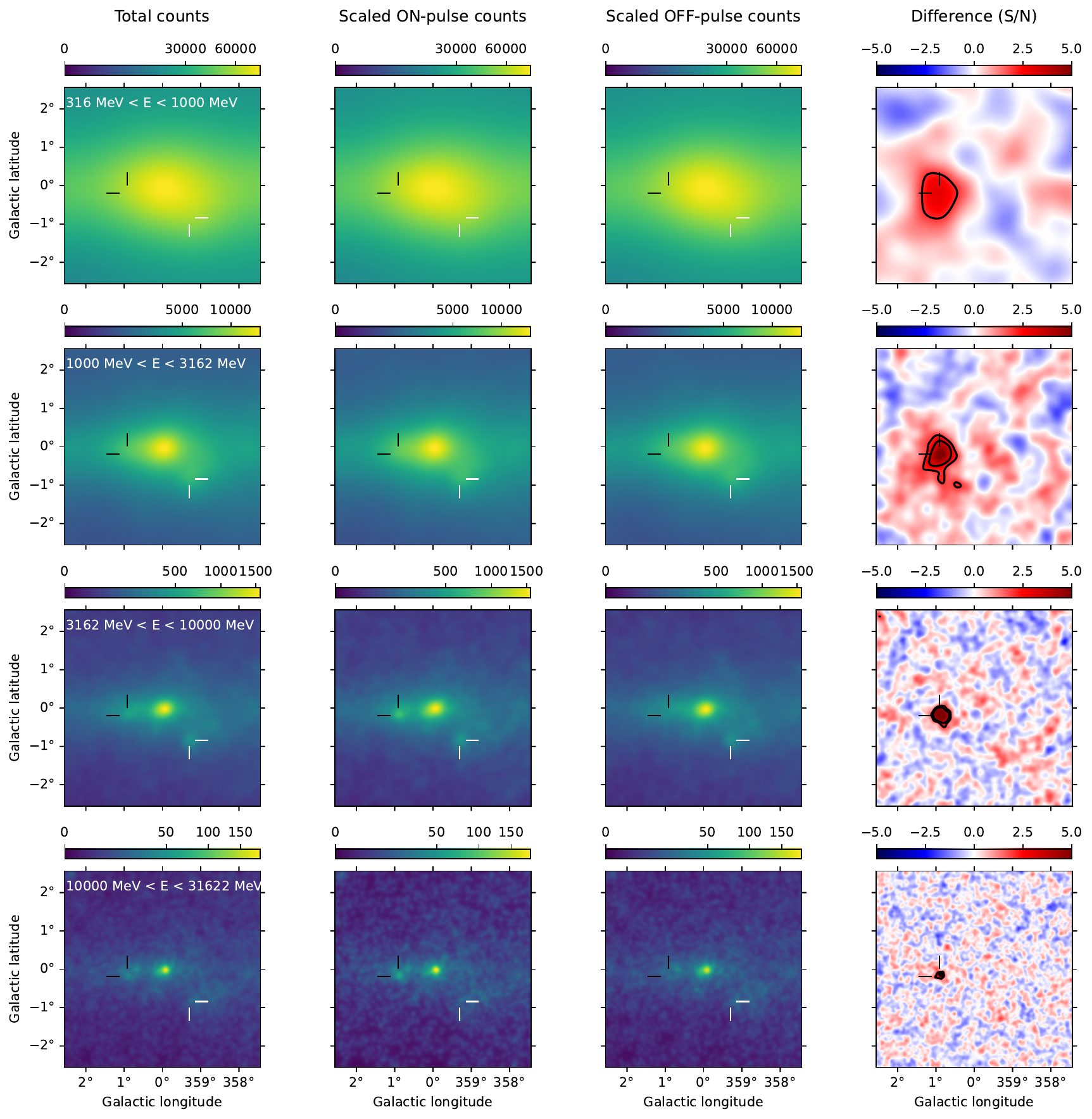}
\caption{\label{f:J1748-2815_diff} Gamma-ray photon counts maps of the Galactic
  Center region. Each row shows counts maps for logarithmically-spaced energy
  bands, convolved with the \textit{Fermi}-LAT point-spread function for the
  highest energy in that band. The gamma-ray timing position of PSR~J1748$-$2815
  is highlighted by black markers. In each row panels show, from left to right: the
  total photon counts, on-pulse photon counts (divided by the on-pulse phase fraction
  to maintain the same color scale as the total count maps), off-pulse photon counts
  (divided by the off-pulse phase fraction), and the difference between the rescaled on- and
  off-pulse maps in signal-to-noise units. Counts maps have a square-root scaling, 
  indicated on the color bars. Contour lines on the difference
  images are at the $2\sigma$ and $3\sigma$ level. The point source visible at
  $(l,b) = (359.3^{\circ},-0.8^{\circ})$, highlighted by white markers, is another gamma-ray pulsar,
  PSR~J1747$-$2958.}
\end{figure*}

Assuming there is no un-pulsed emission from this pulsar, which is the case for most known gamma-ray pulsars \citepalias{3PC}, the on-pulse map
selection should contain all of the flux from the pulsar, while it should be
entirely absent from the off-pulse map. Subtracting the off-pulse counts from
the on-pulse counts (weighted by the relative phase ranges covered by the on-
and off-pulse phase selections) should therefore result in a difference image
where the background is consistent with zero (within the uncertainties), and
remaining excess flux is entirely attributable to the pulsar. We can therefore
fit the spectrum of the excess counts to estimate the spectrum of the pulsar,
without needing to model the flux of any other source, including the GC, and the
diffuse interstellar emission. The on- and off-pulse counts maps and difference
images for different energy bands are shown in
Figure~\ref{f:J1748-2815_diff}. The Mouse pulsar, J1747$-$2958, is also visible
as a point source in these counts maps. Its spectrum could be similarly
estimated from the on- and off-pulse counts to aid investigations the Galactic
Center flux, although its larger separation from the Galactic plane ($|b| = 0.84^{\circ}$)
makes this less crucial.

We modeled the pulsar spectrum as a sub-exponentially-cutoff power law, $dN/dE
\propto (E/1000~{\rm MeV})^{-\Gamma} \exp(-aE^{2/3})$ and used the Fermi tools 
to compute the expected count map from a point source at the pulsar position. 
Fitting the count spectrum within the PSF 68\% containment radius gives a 
spectral index $\Gamma = 1.07 \pm 0.16 \pm 0.09$, exponential factor $a=(6.0 \pm 1.0 \pm 0.03)\times10^{-3}$, and
an integrated energy flux above 100~MeV of $(2.8 \pm 0.2 \pm 0.4)\times10^{-11}~{\rm
  erg}~{\rm cm}^{-2}~{\rm s}^{-1}$, where the first (statistical) uncertainties are obtained using the wider on-pulse phase definition (0.26--0.65), and the second (systematic) uncertainties show the shift from changing to the narrower (0.34--0.65) on-pulse region. This flux is only around 60\% of the value
  for this source in \citet{3FGL}, suggesting that the previous catalog analyses
  significantly over-estimated the flux from this source.

Comparing this energy flux to the pulsar's spin-down luminosity, $\dot{E} =
1.4\times10^{35}~{\rm erg}~{\rm s}^{-1}$, we find a 100\%-efficiency distance
upper limit of around $7$~kpc, but a heuristic luminosity distance of $1~{\rm
  kpc} < d < 4~{\rm kpc}$. It is therefore most likely that this is also a
foreground pulsar, rather than one that originated from a GC pulsar population.

The timing noise model for this pulsar has a lower amplitude than that of
PSR~J1736$-$3422, and is consistent with a simple power-law form with $\gamma >
2.5$. These properties are broadly
consistent with those of similarly energetic gamma-ray pulsars timed by
\citet{Kerr2015+Timing}.

We found archival radio observations from Murriyang covering the location of PSR
J1748$-$2815 as part of the Breakthrough Listen project
\citep{Gajjar2021+BLGC,PX600}, using the Ultra-Wide-bandwidth Low-frequency
(UWL) receiver \citep{UWL}. These consisted of observations on two epochs in
which this position was covered three times with 10-minute pointings separated
by 10-minute observations of another nearby field. We found no significant radio
pulsations in any individual 10-minute observation, or in the stacked
data. While our gamma-ray ephemeris is sufficient for stacking observations
taken on the same epoch, the uncertainty on our timing solution is large enough
that there may be a small relative phase offset between the two observations,
greater than the phase binning of our radio fold, meaning that we cannot
guarantee that a putative radio pulse would lie within the same phase bin in
both observations. We therefore report upper limits on the radio flux density
from single epochs, rather than the upper limit from stacking all of the
Murriyang data, although this also resulted in a non-detection.

The MeerKAT Galactic Center Mosaic \citep{Heywood2022+MKGC}, shown in
Figure~\ref{f:radio_images}, shows that PSR~J1748$-$2815 lies close to the SNR
G1.0$-$0.1, near the end of a tail-like structure protruding from the SNR, and
also near the end of a likely non-thermal radio filament, although the gamma-ray
timing position is significantly offset from all of these. Given the high
density of radio-emitting sources in the busy GC region, and that our heuristic
distance estimate places this pulsar well in front of the GC, it seems unlikely
that the pulsar is associated with any of these structures.

\section{Discussion}
\label{s:discussion}
We were motivated to perform our survey for new gamma-ray pulsars in the inner
Galaxy to identify bright, isolated MSPs that from a putative
bulge population that could explain the observed Galactic center gamma-ray excess.

We found in Section~\ref{s:psrs} that all of the new pulsars are likely to lie
significantly closer to Earth than to the GC, making it likely that they all 
belong to the standard Galactic disk population. However, for the one new MSP 
that we discovered, PSR~J1649$-$3012, there was a chance that it could lie at 
larger distances from Earth, up to around 8~kpc, that could make an origin in 
the Galactic bulge possible. To evaluate the chance of this MSP being from the 
putative bulge population, we
compared the predicted MSP densities along the line-of-sight to PSR~J1649$-$3012
from models of the Galactic thick-disk and bulge MSP populations. For the
former, we took the model from \citet{Levin2013+GalMSPs}, which has a total
of $83000\pm42000$ MSPs with a Gaussian distribution with width $4.5~$kpc in
cylindrical radius around the Galactic center and an exponential distribution in
the height above the Galactic plane with scale height $0.5$~kpc. For the bulge
population, we took the model from \citet{Calore2016}, which has a total of
$9200\pm3100$ MSPs within a radius of 3\,kpc around the center of the Galaxy 
(similar to our $40^{\circ} \times 40^{\circ}$ search region), in
a spherically symmetrical distribution with density falling off with
galactocentric distance as $r^{-2.56}$. The ratio of these two densities
$\rho_{\rm bulge}/\rho_{\rm disk}$ is shown in Figure~\ref{f:J1649_dist}. The
disk model predicts a larger MSP density than the bulge model at all distances,
by more than an order-of-magnitude for distances below 6\,kpc, but this ratio
could be close to unity at distances of around 8\,kpc. Hence it is not
impossible that this MSP originated in the Galactic bulge, but given that a wide
range of more typical efficiencies result in distances for which the disk model
is strongly favored, we conclude that it is more likely that this pulsar
belongs to the normal Galactic disk population.

However, while we find that it is unlikely that we have found any new pulsars 
belonging to a putative bulge population, we emphasize that our results in no
way rule out the presence of such a population. 

Due to the necessary compromise between computing cost and sensitivity, our
semi-coherent search method is only able to detect gamma-ray pulsations from
fairly bright sources, and could therefore only identify new GC MSPs in the
upper tail of the luminosity distribution. To estimate how much less sensitive
our searches are compared to a fully-coherent search for pulsations from a known
pulsar where the ephemeris parameters are already measured from a radio timing
campaign, we consider the relevant test statistics, $S$ and $\mathcal{P}$. As
derived in \citet{Nieder2020+Methods}, both test statistics have spectral
signal-to-noise ratios $\theta^2_{S}$ and $\theta^2_{\mathcal{P}}$ proportional
to the sum of the squared photon weights ($W^2 = \sum_j w_j^2$), with
$\theta^2_{S}$ being lower by a factor of $\sqrt{T/T_{\rm obs}}$. We show in
Appendix~\ref{a:W2} that $W^2$ is proportional to the square of the pulsar
photon flux.

These test statistics have different noise distributions (normal and
chi-squared, respectively) and the two search methods have different associated
trials factors, meaning we must consider different significance thresholds for
each. \citet{Smith2019+ThousandPSRs} showed that a threshold of $H=25$ is a
sufficiently conservative threshold for detecting gamma-ray pulsations when
folding with a radio ephemeris. Since most gamma-ray pulse profiles have most
power in the fundamental harmonic, we consider a threshold of $\mathcal{P}
\approx 20$, corresponding to $\theta^2_{\mathcal{P}} = 9$. For our
semi-coherent searches, we find an empirical threshold of $\theta^2_{S} = 8$ for
candidate signals entering the follow-up stages of our hierarchical search.

We can therefore estimate the relative flux thresholds of our
semi-coherent search compared to searches for gamma-ray pulsations from radio
pulsars with well-known ephemerides as:
\begin{equation}
\begin{aligned}
\frac{F^{\ast}_{S}}{F^{\ast}_{\mathcal{P}}} &= \left(\frac{\theta^2_{S}}{\theta^2_{\mathcal{P}}}\right)^{1/2} \left(\frac{T_{\rm obs}}{T}\right)^{1/4}\,, \\
&= \left(\frac{8}{9}\right)^{1/2} \left(\frac{16\,{\rm yr}}{2^{22}\,{\rm s}}\right)\,,\\
&\approx 3.1\,.
\end{aligned}
\end{equation}
\citet{Smith2019+ThousandPSRs} demonstrated that gamma-ray pulsations can be
detected from sources all the way down to, and sometimes even below, the
\citetalias{4FGL} point-source threshold. Hence, the coherent Fourier power has
a similar photon flux threshold as the \textit{Fermi}-LAT point-source
significance threshold. We are therefore only sensitive to sources
that have photon fluxes that are around 3.1 times larger than the 4FGL detection
threshold to be detectable in our searches.

Recent analyses of the GC excess have found that, under the MSP population
explanation, only $\mathcal{O}(20)$ MSPs should be individually detectable as
gamma-ray point sources in the latest \textit{Fermi}-LAT data
\citep[e.g.][]{Holst2024,Malyshev2024}. The relatively higher threshold of our
semi-coherent search means that only around 50\% of these are bright enough to
be detectable in our searches, leaving only around $\mathcal{O}(10)$ detectable pulsars. More importantly, around 80\% of Galactic MSPs
are in binary systems, and therefore undetectable to our searches, and this
fraction may even be under-estimated due to historic biases against detecting
short-period binary MSPs \citep{Bangale2024}.

The number of GC MSPs that we might have expected to detect in a perfect survey 
for isolated pulsars in inner-Galaxy sources is therefore likely to be only around
one or two. The fact that we did not detect any MSPs that can be attributed to 
this population is therefore not surprising, nor at odds with the MSP explanation 
for the GC excess.

The lack of sensitivity of our searches to binary MSPs may indeed be one of the
largest barriers to detecting the GC MSP population through gamma-ray pulsation
searches. The new pulsation searching method of \citet{Gazith2024}
does not have this shortcoming, although it remains unclear
how well that method will cope in the inner Galaxy, where the high background
level means that pulsar signals tend to consist of a large number of relatively
low-weight photons, rather than a smaller number of high-weight photons.

The lack of candidate GC MSPs in \citetalias{3PC} has been interpreted as
evidence against the GC population \citep{Holst2024}. However, the detection of
a pulsar as a gamma-ray point source was not a sufficient (nor even necessary)
criterion for inclusion in \citetalias{3PC}, where the only requirement was the
significant detection of gamma-ray pulsations. For faint sources, such a
detection requires a radio timing solution that either covers or can be
extrapolated back to the start of the \textit{Fermi} mission. Radio pulsar
searches of the inner Galaxy are, however, particularly challenging, due to the
large distance and high background of radio-emitting sources that reduce
sensitivity, and the high level of scattering and dispersion from the
interstellar medium that smear out pulses and particularly affect searches for
rapidly spinning MSPs. These propagation effects are particularly strong at low
radio frequencies $\lesssim \mathcal{O}(1\,{\rm GHz})$, where most large-scale
pulsar surveys have historically taken place, and where pulsars tend to be
brightest due to their steep power-law spectra \citep{Karastergiou2024+TPA}.
The particular challenge of detecting MSPs in the inner Galaxy is highlighted by
the fact that the only known MSP within 1$^{\circ}$ of the GC was only recently
discovered \citep{Lower2024}, and was only detectable at $\gtrsim 2~{\rm GHz}$,
requiring the novel wide-band high-frequency coverage of the UWL receiver on
Murriyang \citep{UWL}. 

Recent analyses have found that some sources in the inner Galaxy are now
becoming individually detectable as resolved sources, with spectra consistent
with MSPs \citep{Malyshev2024}, but it may require next-generation radio
telescopes, and long radio timing campaigns before these can be identified as
gamma-ray pulsars. 

Due to the large overlap between the expected bulge and disk populations 
(evident in the lower panel of Figure 3), evidence for a bulge MSP population that could explain the GC gamma-ray excess
will require the detection of a large overdensity of MSPs in the inner Galaxy 
relative to the rest of the Galactic disk. Efforts are underway to identify 
candidate pulsars in X-ray \citep{Berteaud2021,Berteaud2024} and radio imaging 
surveys \citep{Frail2024}, and indeed three new MSPs have recently been discovered by targeting 
candidate sources identified by these surveys \citep[J. Berteaud, A\&A, submitted;][]{Hyman2025}. 
These may represent the tip of the iceberg for the bulge population.  
While we did not find any new MSPs here that are likely to be 
associated to the bulge population, we have shown that 
gamma-ray searches could possibly contribute one or two new members 
to this population. A renewed survey similar to the one performed here, 
but with a longer data set, new gamma-ray sources, and a longer coherence time to 
increase sensitivity, may yet contribute to uncovering this population.

\section{Conclusions}
\label{s:conc}
We performed searches for gamma-ray pulsations from 55 sources in the inner
Galaxy ($-20^{\circ} < l < 20^{\circ}$, $-20^{\circ} < b < 20^{\circ}$),
identifying 4 of these as new gamma-ray pulsars. None of these pulsars
have been detected in radio observations, including the isolated MSP
J1649$-$3012 that is now only the third confirmed ``radio-quiet'' gamma-ray
MSP.

One of these pulsars, J1748$-$2815, has the smallest angular separation to the
GC of any known gamma-ray pulsar, at $0.93^{\circ}$. This pulsar is not detected
as a point source in \citetalias{4FGL-DR4}, likely due to some of its flux being
misattributed to the diffuse background in that direction. This suggests that a 
strategy of searching a large grid of sky locations covering the innermost part 
of the Galaxy, using the methods of \citet{Bruel2019} to weight photons regardless
of whether or not a significant source is detected at that position, may be able to 
reveal additional pulsars. We performed a
phase-resolved analysis that allows us to disentangle its flux from the
surrounding area, which provides a spectral anchor point for the
development of new interstellar gamma-ray emission models.

The estimated distances to these pulsars, based on their measured gamma-ray
fluxes and assuming typical spin-down/gamma-ray efficiencies suggest that they
are all more likely to be foreground sources than from a putative bulge
population that may explain the GC GeV excess, but our results in no way rule
out the presence of such a population.

Nevertheless, our discoveries demonstrate our ability to find new pulsars in
this region of the sky, and suggest that new searches of the inner Galaxy, with
the extra years of LAT data and newly identified gamma-ray sources that have
accumulated since we began this project, may soon be able to identify the brightest
isolated pulsars from this putative population.

\section*{acknowledgments}
We gratefully acknowledge the contributions of the \textit{Einstein@Home} 
volunteers, without whom this work would not have been possible. We would particularly 
like to thank the volunteers whose computers discovered the new pulsars presented in this work,
\begin{itemize}
    \item PSR~J1649$-$3012: Alexander Kravchenko, Moscow, Russia, and the ATLAS cluster, Hannover, Germany; 
    \item PSR~J1736$-$3422: the ATLAS cluster, Hannover, Germany; 
    \item PSR~J1742$-$3321: Lingnan Cai, Nanjing, China, and the ATLAS cluster, Hannover, Germany; 
    \item PSR~J1748$-$2815: Duncan Brown, Syracuse, NY, USA, and Miguel Angel Gonzalo, Madrid, Spain. 
\end{itemize}
\textit{Einstein@Home} is supported by National Science
Foundation grant NSF~1816904.

Work at NRL is supported by NASA.

We thank Paul S. Ray and Giacomo Principe for reviewing this paper on behalf of the \textit{Fermi}-LAT 
collaboration. We also thank Bhaswati Bhattacharyya and Jayanta Roy for investigating 
archival GMRT observations to check for previous coverage of PSR~J1649$-$3012. 

The \textit{Fermi} LAT Collaboration acknowledges generous ongoing support from
a number of agencies and institutes that have supported both the development and
the operation of the LAT as well as scientific data analysis.  These include the
National Aeronautics and Space Administration and the Department of Energy in
the United States, the Commissariat \`a l'Energie Atomique and the Centre
National de la Recherche Scientifique / Institut National de Physique
Nucl\'eaire et de Physique des Particules in France, the Agenzia Spaziale
Italiana and the Istituto Nazionale di Fisica Nucleare in Italy, the Ministry of
Education, Culture, Sports, Science and Technology (MEXT), High Energy
Accelerator Research Organization (KEK) and Japan Aerospace Exploration Agency
(JAXA) in Japan, and the K.~A.~Wallenberg Foundation, the Swedish Research
Council and the Swedish National Space Board in Sweden.
Additional support for science analysis during the operations phase is gratefully
acknowledged from the Istituto Nazionale di Astrofisica in Italy and the Centre
National d'\'Etudes Spatiales in France. This work performed in part under DOE
Contract DE-AC02-76SF00515.

The MeerKAT telescope is operated by the South African Radio Astronomy Observatory, which is a facility of the National Research Foundation, an agency of the Department of Science and Innovation. This work has made use of the “MPIfR S-band receiver system” designed, constructed and maintained by funding of the MPI f\"{u}r Radioastronomie (MPIfR) and the Max-Planck-Society (MPG). Observations used the FBFUSE and APSUSE computing clusters for data acquisition, storage and analysis. These clusters were funded, designed and installed by the MPIfR and the MPG. FBFUSE performs beamforming operations in real-time using the \texttt{mosaic}\footnote{\url{https://github.com/wchenastro/Mosaic}} software stack \citep{FBFUSE}. The MPIfR-MeerKAT Galactic Plane Survey is led by the MPIfR and is supported by the MPG.

Murriyang, CSIRO’s Parkes radio telescope, is part of the Australia Telescope National Facility\footnote{\url{https://ror.org/05qajvd42}} which is funded by the Australian Government for operation as a National Facility managed by CSIRO. We acknowledge the Wiradjuri people as the Traditional Owners of the Observatory site.

The Green Bank Observatory is a facility of the National Science Foundation operated under cooperative agreement by Associated Universities, Inc.

This research has made use of data or software obtained from the Gravitational Wave Open Science Center (gwosc.org), a service of the LIGO Scientific Collaboration, the Virgo Collaboration, and KAGRA. This material is based upon work supported by NSF's LIGO Laboratory which is a major facility fully funded by the National Science Foundation, as well as the Science and Technology Facilities Council (STFC) of the United Kingdom, the Max-Planck-Society (MPS), and the State of Niedersachsen/Germany for support of the construction of Advanced LIGO and construction and operation of the GEO600 detector. Additional support for Advanced LIGO was provided by the Australian Research Council. Virgo is funded, through the European Gravitational Observatory (EGO), by the French Centre National de Recherche Scientifique (CNRS), the Italian Istituto Nazionale di Fisica Nucleare (INFN) and the Dutch Nikhef, with contributions by institutions from Belgium, Germany, Greece, Hungary, Ireland, Japan, Monaco, Poland, Portugal, Spain. KAGRA is supported by Ministry of Education, Culture, Sports, Science and Technology (MEXT), Japan Society for the Promotion of Science (JSPS) in Japan; National Research Foundation (NRF) and Ministry of Science and ICT (MSIT) in Korea; Academia Sinica (AS) and National Science and Technology Council (NSTC) in Taiwan. 

\facilities{Fermi, GBT, LIGO, MeerKAT, Parkes}
\software{Astropy \citep{astropy:2013, astropy:2018, astropy:2022}, bilby \citep{Bilby}, BOINC \citep{BOINC}, dspsr \citep{dspsr}, dynesty \citep{dynesty,dynestypy},  Fermitools \citep{fermitools}, fermipy \citep{fermipy}, FFTW \citep{FFTW}, GalPot \citep{GalPot}, LALSuite \citep{LALSuite}, matplotlib \citep{matplotlib}, numpy \citep{numpy}, psrchive \citep{psrchive+software}, PulsarX \citep{PulsarX}, PINT \citep{Luo2018+PINT}}

\bibliographystyle{aasjournalv7}
\bibliography{ms}

\appendix
\restartappendixnumbering
\section{List of targeted sources}
\label{a:sources}

    \startlongtable
    \begin{deluxetable*}{lcccccccc}
    \tabletypesize{\scriptsize}
\tablecaption{\label{t:sources}List of sources searched for pulsations by Einstein@Home}
\tablehead{
    \colhead{4FGL Source} &
    \colhead{2FIG Source} &
    \colhead{R.A.} &
    \colhead{Decl.} &
    \colhead{$l$ $\left(\degr\right)$} &
    \colhead{$b$ $\left(\degr\right)$} &
    \colhead{$r_{\rm 95\%}$ $\left(\arcmin\right)$ \tablenotemark{*}} &
    \colhead{$p_{\rm min}$ \tablenotemark{$\dagger$}} &
    \colhead{Spectral models \tablenotemark{$\ddagger$}}
    }
\startdata
    \hline
 J1624.3-3952\tablenotemark{b} & J1624.4-3952 &  $16^{\mathrm{h}}24^{\mathrm{m}}29^{\mathrm{s}}$ & $-39^\circ52{}^\prime45{}^{\prime\prime}$ & $-18.812$ & $6.670$ & $2.32$ & $0.864$ & $>100$\,MeV, both (Off.)\\
 & &  $16^{\mathrm{h}}24^{\mathrm{m}}29^{\mathrm{s}}$ & $-39^\circ52{}^\prime45{}^{\prime\prime}$ & $-18.812$ & $6.670$ & $2.33$ & $0.898$ & $>300$\,MeV, both (Off.)\\
\hline
 --- & J1627.8-2436 &  $16^{\mathrm{h}}27^{\mathrm{m}}51^{\mathrm{s}}$ & $-24^\circ36{}^\prime52{}^{\prime\prime}$ & $-6.874$ & $16.527$ & $4.86$ & $0.595$ & $>100$\,MeV, both (Alt.)\\
\hline
 J1641.3-2908 & J1641.3-2907 &  $16^{\mathrm{h}}41^{\mathrm{m}}23^{\mathrm{s}}$ & $-29^\circ07{}^\prime47{}^{\prime\prime}$ & $-8.376$ & $11.296$ & $6.43$ & $0.717$ & $>100$\,MeV, both (Alt.)\\
\hline
 J1649.3-4441 & J1649.4-4440 &  $16^{\mathrm{h}}49^{\mathrm{m}}26^{\mathrm{s}}$ & $-44^\circ40{}^\prime20{}^{\prime\prime}$ & $-19.281$ & $0.038$ & $2.86$ & $0.679$ & $>100$\,MeV, Off. only\\
\hline
 J1649.8-3010\tablenotemark{a} & J1649.8-3010 &  $16^{\mathrm{h}}49^{\mathrm{m}}51^{\mathrm{s}}$ & $-30^\circ10{}^\prime29{}^{\prime\prime}$ & $-8.007$ & $9.213$ & $2.01$ & $0.676$ & $>100$\,MeV, Off. only\\
\hline
 J1650.9-4420c & J1651.3-4417 &  $16^{\mathrm{h}}51^{\mathrm{m}}18^{\mathrm{s}}$ & $-44^\circ17{}^\prime27{}^{\prime\prime}$ & $-18.774$ & $0.025$ & $5.24$ & $0.544$ & $>100$\,MeV, Off. only\\
\hline
 --- & J1652.1-4430 &  $16^{\mathrm{h}}52^{\mathrm{m}}07^{\mathrm{s}}$ & $-44^\circ30{}^\prime05{}^{\prime\prime}$ & $-18.845$ & $-0.220$ & $5.59$ & $0.895$ & $>300$\,MeV, both (Off.)\\
\hline
 --- & J1652.7-4352 &  $16^{\mathrm{h}}52^{\mathrm{m}}47^{\mathrm{s}}$ & $-43^\circ52{}^\prime50{}^{\prime\prime}$ & $-18.288$ & $0.080$ & $2.56$ & $0.525$ & $>100$\,MeV, both (Off.)\\
 & &  $16^{\mathrm{h}}52^{\mathrm{m}}47^{\mathrm{s}}$ & $-43^\circ52{}^\prime50{}^{\prime\prime}$ & $-18.288$ & $0.080$ & $2.29$ & $0.455$ & $>300$\,MeV, both (Off.)\\
\hline
 J1653.2-4349 & J1652.8-4351 &  $16^{\mathrm{h}}53^{\mathrm{m}}12^{\mathrm{s}}$ & $-43^\circ49{}^\prime27{}^{\prime\prime}$ & $-18.197$ & $0.058$ & $2.73$ & $0.57$ & 4FGL\\
\hline
 J1707.1-1931\tablenotemark{c} & J1706.9-1932 &  $17^{\mathrm{h}}07^{\mathrm{m}}00^{\mathrm{s}}$ & $-19^\circ32{}^\prime23{}^{\prime\prime}$ & $3.021$ & $12.513$ & $3.99$ & $0.646$ & $>100$\,MeV, Off. only\\
\hline
 J1710.3-3943\tablenotemark{c} & J1709.8-3944 &  $17^{\mathrm{h}}09^{\mathrm{m}}54^{\mathrm{s}}$ & $-39^\circ44{}^\prime12{}^{\prime\prime}$ & $-13.061$ & $0.109$ & $2.39$ & $0.75$ & $>300$\,MeV, both (Off.)\\
\hline
 J1709.9-0900 & J1710.0-0905 &  $17^{\mathrm{h}}10^{\mathrm{m}}06^{\mathrm{s}}$ & $-09^\circ05{}^\prime24{}^{\prime\prime}$ & $12.454$ & $17.700$ & $7.79$ & $0.846$ & $>100$\,MeV, both (Off.)\\
\hline
 J1711.0-3002 & J1711.1-3004 &  $17^{\mathrm{h}}11^{\mathrm{m}}08^{\mathrm{s}}$ & $-30^\circ04{}^\prime52{}^{\prime\prime}$ & $-5.113$ & $5.616$ & $4.32$ & $0.868$ & $>300$\,MeV, both (Off.)\\
\hline
 J1714.9-3324 & J1715.0-3324 &  $17^{\mathrm{h}}15^{\mathrm{m}}01^{\mathrm{s}}$ & $-33^\circ24{}^\prime34{}^{\prime\prime}$ & $-7.338$ & $3.001$ & $3.48$ & $0.985$ & $>100$\,MeV, both (Off.)\\
\hline
 J1720.6-3706\tablenotemark{e} & J1720.4-3707 &  $17^{\mathrm{h}}20^{\mathrm{m}}26^{\mathrm{s}}$ & $-37^\circ07{}^\prime25{}^{\prime\prime}$ & $-9.736$ & $-0.044$ & $5.71$ & $0.82$ & $>100$\,MeV, Off. only\\
\hline
 J1729.1-3503 & J1729.1-3502 &  $17^{\mathrm{h}}29^{\mathrm{m}}11^{\mathrm{s}}$ & $-35^\circ02{}^\prime39{}^{\prime\prime}$ & $-7.023$ & $-0.337$ & $3.59$ & $0.915$ & $>100$\,MeV, both (Off.)\\
\hline
 J1730.4-0359\tablenotemark{b} & J1730.3-0400 &  $17^{\mathrm{h}}30^{\mathrm{m}}23^{\mathrm{s}}$ & $-04^\circ00{}^\prime21{}^{\prime\prime}$ & $19.719$ & $16.000$ & $2.11$ & $0.616$ & $>100$\,MeV, both (Off.)\\
 & &  $17^{\mathrm{h}}30^{\mathrm{m}}23^{\mathrm{s}}$ & $-04^\circ00{}^\prime21{}^{\prime\prime}$ & $19.719$ & $16.000$ & $2.07$ & $0.603$ & $>300$\,MeV, both (Off.)\\
 & &  $17^{\mathrm{h}}30^{\mathrm{m}}26^{\mathrm{s}}$ & $-03^\circ59{}^\prime32{}^{\prime\prime}$ & $19.738$ & $15.995$ & $2.21$ & $0.65$ & 4FGL\\
\hline
 J1731.6-3002\tablenotemark{c} & J1731.6-3001 &  $17^{\mathrm{h}}31^{\mathrm{m}}41^{\mathrm{s}}$ & $-30^\circ02{}^\prime45{}^{\prime\prime}$ & $-2.558$ & $1.973$ & $2.43$ & $0.794$ & 4FGL\\
\hline
 --- & J1732.9-2904 &  $17^{\mathrm{h}}33^{\mathrm{m}}00^{\mathrm{s}}$ & $-29^\circ04{}^\prime48{}^{\prime\prime}$ & $-1.593$ & $2.261$ & $5.73$ & $0.533$ & $>300$\,MeV, both (Alt.)\\
\hline
 --- & J1734.5-3239 &  $17^{\mathrm{h}}34^{\mathrm{m}}35^{\mathrm{s}}$ & $-32^\circ39{}^\prime04{}^{\prime\prime}$ & $-4.406$ & $0.034$ & $3.18$ & $0.679$ & $>100$\,MeV, Off. only\\
\hline
 --- & J1735.5-3218 &  $17^{\mathrm{h}}35^{\mathrm{m}}34^{\mathrm{s}}$ & $-32^\circ18{}^\prime38{}^{\prime\prime}$ & $-4.007$ & $0.043$ & $2.95$ & $0.925$ & $>100$\,MeV, Off. only\\
\hline
 J1736.1-3422\tablenotemark{a} & J1736.2-3422 &  $17^{\mathrm{h}}36^{\mathrm{m}}06^{\mathrm{s}}$ & $-34^\circ22{}^\prime38{}^{\prime\prime}$ & $-5.687$ & $-1.166$ & $2.57$ & $0.581$ & 4FGL\\
\hline
 J1738.2-2510 & J1738.3-2509 &  $17^{\mathrm{h}}38^{\mathrm{m}}18^{\mathrm{s}}$ & $-25^\circ09{}^\prime19{}^{\prime\prime}$ & $2.356$ & $3.377$ & $7.55$ & $0.345$ & $>100$\,MeV, both (Off.)\\
\hline
 --- & J1740.4-2908 &  $17^{\mathrm{h}}40^{\mathrm{m}}26^{\mathrm{s}}$ & $-29^\circ08{}^\prime56{}^{\prime\prime}$ & $-0.776$ & $0.851$ & $4.35$ & $0.911$ & $>100$\,MeV, Off. only\\
\hline
 J1740.7-2640 & J1740.5-2641 &  $17^{\mathrm{h}}40^{\mathrm{m}}32^{\mathrm{s}}$ & $-26^\circ41{}^\prime08{}^{\prime\prime}$ & $1.324$ & $2.140$ & $5.29$ & $0.448$ & $>100$\,MeV, Off. only\\
\hline
 J1740.4-2850 & J1740.6-2845 &  $17^{\mathrm{h}}40^{\mathrm{m}}37^{\mathrm{s}}$ & $-28^\circ45{}^\prime17{}^{\prime\prime}$ & $-0.421$ & $1.027$ & $3.04$ & $0.779$ & $>100$\,MeV, Off. only\\
\hline
 J1742.3-3318\tablenotemark{a} & J1742.2-3318 &  $17^{\mathrm{h}}42^{\mathrm{m}}17^{\mathrm{s}}$ & $-33^\circ18{}^\prime46{}^{\prime\prime}$ & $-4.103$ & $-1.684$ & $2.33$ & $0.444$ & $>100$\,MeV, both (Off.)\\
\hline
 J1743.9-1824\tablenotemark{c} & J1743.3-1827 &  $17^{\mathrm{h}}43^{\mathrm{m}}22^{\mathrm{s}}$ & $-18^\circ27{}^\prime16{}^{\prime\prime}$ & $8.697$ & $5.896$ & $7.20$ & $0.713$ & $>100$\,MeV, Off. only\\
\hline
 J1744.0-1311 & J1744.0-1311 &  $17^{\mathrm{h}}44^{\mathrm{m}}06^{\mathrm{s}}$ & $-13^\circ11{}^\prime33{}^{\prime\prime}$ & $13.339$ & $8.447$ & $2.51$ & $0.96$ & $>100$\,MeV, both (Off.)\\
 & &  $17^{\mathrm{h}}44^{\mathrm{m}}06^{\mathrm{s}}$ & $-13^\circ11{}^\prime33{}^{\prime\prime}$ & $13.339$ & $8.447$ & $2.56$ & $0.965$ & $>300$\,MeV, both (Off.)\\
\hline
 J1744.7-1557 & J1744.7-1557 &  $17^{\mathrm{h}}44^{\mathrm{m}}46^{\mathrm{s}}$ & $-15^\circ57{}^\prime22{}^{\prime\prime}$ & $11.024$ & $6.898$ & $3.68$ & $0.981$ & $>300$\,MeV, both (Off.)\\
\hline
 --- & J1747.8-3013 &  $17^{\mathrm{h}}47^{\mathrm{m}}50^{\mathrm{s}}$ & $-30^\circ13{}^\prime10{}^{\prime\prime}$ & $-0.848$ & $-1.076$ & $2.39$ & $0.366$ & $>300$\,MeV, both (Alt.)\\
\hline
 --- & J1748.4-2814\tablenotemark{a} &  $17^{\mathrm{h}}48^{\mathrm{m}}28^{\mathrm{s}}$ & $-28^\circ14{}^\prime33{}^{\prime\prime}$ & $0.918$ & $-0.176$ & $0.90$ & $0.36$ & $>300$\,MeV, both (Alt.)\\
\hline
 J1748.8-3915 & J1748.6-3912 &  $17^{\mathrm{h}}48^{\mathrm{m}}42^{\mathrm{s}}$ & $-39^\circ12{}^\prime46{}^{\prime\prime}$ & $-8.491$ & $-5.845$ & $4.06$ & $0.65$ & $>100$\,MeV, both (Off.)\\
 & &  $17^{\mathrm{h}}48^{\mathrm{m}}42^{\mathrm{s}}$ & $-39^\circ12{}^\prime46{}^{\prime\prime}$ & $-8.491$ & $-5.845$ & $3.78$ & $0.624$ & $>300$\,MeV, both (Off.)\\
\hline
 --- & J1753.3-4446 &  $17^{\mathrm{h}}53^{\mathrm{m}}19^{\mathrm{s}}$ & $-44^\circ46{}^\prime56{}^{\prime\prime}$ & $-12.932$ & $-9.364$ & $2.44$ & $0.67$ & $>100$\,MeV, both (Off.)\\
 & &  $17^{\mathrm{h}}53^{\mathrm{m}}19^{\mathrm{s}}$ & $-44^\circ46{}^\prime56{}^{\prime\prime}$ & $-12.932$ & $-9.364$ & $2.47$ & $0.632$ & $>300$\,MeV, both (Off.)\\
\hline
 J1753.8-2538 & J1753.9-2538 &  $17^{\mathrm{h}}53^{\mathrm{m}}50^{\mathrm{s}}$ & $-25^\circ38{}^\prime25{}^{\prime\prime}$ & $3.766$ & $0.127$ & $0.95$ & $0.344$ & 4FGL\\
 & &  $17^{\mathrm{h}}53^{\mathrm{m}}55^{\mathrm{s}}$ & $-25^\circ38{}^\prime44{}^{\prime\prime}$ & $3.770$ & $0.109$ & $0.86$ & $0.46$ & $>100$\,MeV, both (Off.)\\
\hline
 --- & J1754.1-2929 &  $17^{\mathrm{h}}54^{\mathrm{m}}08^{\mathrm{s}}$ & $-29^\circ29{}^\prime52{}^{\prime\prime}$ & $0.471$ & $-1.881$ & $2.49$ & $0.415$ & $>300$\,MeV, both (Off.)\\
\hline
 J1754.6-2933 & J1754.1-2930 &  $17^{\mathrm{h}}54^{\mathrm{m}}39^{\mathrm{s}}$ & $-29^\circ33{}^\prime06{}^{\prime\prime}$ & $0.482$ & $-2.007$ & $4.46$ & $0.675$ & 4FGL\\
\hline
 J1758.3-3028\tablenotemark{c} & J1758.3-3028 &  $17^{\mathrm{h}}58^{\mathrm{m}}21^{\mathrm{s}}$ & $-30^\circ28{}^\prime01{}^{\prime\prime}$ & $0.090$ & $-3.157$ & $3.73$ & $0.8$ & $>100$\,MeV, both (Alt.)\\
\hline
 J1758.7-4109 & J1758.8-4108 &  $17^{\mathrm{h}}58^{\mathrm{m}}47^{\mathrm{s}}$ & $-41^\circ09{}^\prime10{}^{\prime\prime}$ & $-9.228$ & $-8.480$ & $2.08$ & $0.878$ & 4FGL\\
 & &  $17^{\mathrm{h}}58^{\mathrm{m}}50^{\mathrm{s}}$ & $-41^\circ09{}^\prime08{}^{\prime\prime}$ & $-9.222$ & $-8.489$ & $2.08$ & $0.816$ & $>100$\,MeV, both (Alt.)\\
 & &  $17^{\mathrm{h}}58^{\mathrm{m}}50^{\mathrm{s}}$ & $-41^\circ09{}^\prime08{}^{\prime\prime}$ & $-9.222$ & $-8.489$ & $2.09$ & $0.917$ & $>300$\,MeV, both (Alt.)\\
\hline
 J1759.1-3849\tablenotemark{c} & J1759.0-3850 &  $17^{\mathrm{h}}59^{\mathrm{m}}10^{\mathrm{s}}$ & $-38^\circ49{}^\prime18{}^{\prime\prime}$ & $-7.131$ & $-7.417$ & $2.48$ & $0.899$ & 4FGL\\
 & &  $17^{\mathrm{h}}59^{\mathrm{m}}07^{\mathrm{s}}$ & $-38^\circ50{}^\prime21{}^{\prime\prime}$ & $-7.153$ & $-7.415$ & $2.34$ & $0.897$ & $>100$\,MeV, both (Off.)\\
 & &  $17^{\mathrm{h}}59^{\mathrm{m}}07^{\mathrm{s}}$ & $-38^\circ50{}^\prime21{}^{\prime\prime}$ & $-7.153$ & $-7.415$ & $2.30$ & $0.906$ & $>300$\,MeV, both (Off.)\\
\hline
 J1759.7-2141\tablenotemark{e} & J1759.6-2141 &  $17^{\mathrm{h}}59^{\mathrm{m}}41^{\mathrm{s}}$ & $-21^\circ41{}^\prime50{}^{\prime\prime}$ & $7.848$ & $0.949$ & $2.07$ & $0.708$ & $>100$\,MeV, both (Off.)\\
\hline
 --- & J1801.5-2248 &  $18^{\mathrm{h}}01^{\mathrm{m}}33^{\mathrm{s}}$ & $-22^\circ48{}^\prime17{}^{\prime\prime}$ & $7.100$ & $0.026$ & $2.53$ & $0.852$ & $>100$\,MeV, Off. only\\
\hline
 J1802.4-3041 & J1802.4-3042 &  $18^{\mathrm{h}}02^{\mathrm{m}}28^{\mathrm{s}}$ & $-30^\circ41{}^\prime57{}^{\prime\prime}$ & $0.326$ & $-4.042$ & $2.78$ & $0.885$ & 4FGL\\
\hline
 J1808.4-3358 & J1808.3-3357 &  $18^{\mathrm{h}}08^{\mathrm{m}}21^{\mathrm{s}}$ & $-33^\circ57{}^\prime32{}^{\prime\prime}$ & $-1.940$ & $-6.704$ & $2.65$ & $0.686$ & $>100$\,MeV, both (Off.)\\
 & &  $18^{\mathrm{h}}08^{\mathrm{m}}21^{\mathrm{s}}$ & $-33^\circ57{}^\prime32{}^{\prime\prime}$ & $-1.940$ & $-6.704$ & $2.63$ & $0.621$ & $>300$\,MeV, both (Off.)\\
 & &  $18^{\mathrm{h}}08^{\mathrm{m}}24^{\mathrm{s}}$ & $-33^\circ58{}^\prime54{}^{\prime\prime}$ & $-1.954$ & $-6.725$ & $2.95$ & $0.602$ & 4FGL\\
\hline
 J1813.2-1128 & J1813.6-1127 &  $18^{\mathrm{h}}13^{\mathrm{m}}37^{\mathrm{s}}$ & $-11^\circ27{}^\prime08{}^{\prime\prime}$ & $18.426$ & $3.025$ & $7.03$ & $0.375$ & $>100$\,MeV, Off. only\\
\hline
 J1819.9-1530 & J1820.2-1526 &  $18^{\mathrm{h}}20^{\mathrm{m}}18^{\mathrm{s}}$ & $-15^\circ26{}^\prime22{}^{\prime\prime}$ & $15.686$ & $-0.297$ & $3.85$ & $0.701$ & $>300$\,MeV, both (Off.)\\
\hline
 J1820.7-3217 & J1820.7-3215 &  $18^{\mathrm{h}}20^{\mathrm{m}}44^{\mathrm{s}}$ & $-32^\circ16{}^\prime00{}^{\prime\prime}$ & $0.785$ & $-8.216$ & $3.65$ & $0.961$ & $>100$\,MeV, both (Off.)\\
\hline
 J1823.3-1340 & J1823.2-1339 &  $18^{\mathrm{h}}23^{\mathrm{m}}17^{\mathrm{s}}$ & $-13^\circ39{}^\prime51{}^{\prime\prime}$ & $17.592$ & $-0.099$ & $0.76$ & $0.287$ & $>300$\,MeV, both (Off.)\\
 & &  $18^{\mathrm{h}}23^{\mathrm{m}}21^{\mathrm{s}}$ & $-13^\circ40{}^\prime03{}^{\prime\prime}$ & $17.596$ & $-0.115$ & $0.98$ & $0.337$ & 4FGL\\
\hline
 J1824.2-5427 & J1824.3-5426 &  $18^{\mathrm{h}}24^{\mathrm{m}}19^{\mathrm{s}}$ & $-54^\circ26{}^\prime54{}^{\prime\prime}$ & $-19.676$ & $-18.053$ & $2.41$ & $0.912$ & $>100$\,MeV, both (Off.)\\
 & &  $18^{\mathrm{h}}24^{\mathrm{m}}19^{\mathrm{s}}$ & $-54^\circ26{}^\prime54{}^{\prime\prime}$ & $-19.676$ & $-18.053$ & $2.36$ & $0.925$ & $>300$\,MeV, both (Off.)\\
\hline
 J1828.0-1133\tablenotemark{d} & J1827.9-1136 &  $18^{\mathrm{h}}27^{\mathrm{m}}56^{\mathrm{s}}$ & $-11^\circ36{}^\prime38{}^{\prime\prime}$ & $19.939$ & $-0.146$ & $5.17$ & $0.688$ & $>100$\,MeV, Off. only\\
\hline
 J1830.8-3132 & J1830.7-3131 &  $18^{\mathrm{h}}30^{\mathrm{m}}47^{\mathrm{s}}$ & $-31^\circ31{}^\prime36{}^{\prime\prime}$ & $2.417$ & $-9.796$ & $2.39$ & $0.989$ & $>100$\,MeV, both (Off.)\\
 & &  $18^{\mathrm{h}}30^{\mathrm{m}}47^{\mathrm{s}}$ & $-31^\circ31{}^\prime36{}^{\prime\prime}$ & $2.417$ & $-9.796$ & $2.36$ & $0.969$ & $>300$\,MeV, both (Off.)\\
\hline
 J1838.9-3457\tablenotemark{c} & J1839.0-3456 &  $18^{\mathrm{h}}39^{\mathrm{m}}03^{\mathrm{s}}$ & $-34^\circ56{}^\prime47{}^{\prime\prime}$ & $0.008$ & $-12.813$ & $3.15$ & $0.972$ & $>100$\,MeV, both (Off.)\\
 & &  $18^{\mathrm{h}}39^{\mathrm{m}}03^{\mathrm{s}}$ & $-34^\circ56{}^\prime47{}^{\prime\prime}$ & $0.008$ & $-12.813$ & $3.09$ & $0.906$ & $>300$\,MeV, both (Off.)\\
\hline
 J1845.8-2521 & J1845.9-2522 &  $18^{\mathrm{h}}45^{\mathrm{m}}52^{\mathrm{s}}$ & $-25^\circ21{}^\prime31{}^{\prime\prime}$ & $9.518$ & $-10.120$ & $3.34$ & $0.902$ & 4FGL\\
\hline
 J1916.8-3025 & J1916.8-3024 &  $19^{\mathrm{h}}16^{\mathrm{m}}52^{\mathrm{s}}$ & $-30^\circ24{}^\prime55{}^{\prime\prime}$ & $7.529$ & $-18.412$ & $1.62$ & $0.694$ & $>100$\,MeV, both (Off.)\\
 & &  $19^{\mathrm{h}}16^{\mathrm{m}}52^{\mathrm{s}}$ & $-30^\circ24{}^\prime55{}^{\prime\prime}$ & $7.529$ & $-18.412$ & $1.61$ & $0.695$ & $>300$\,MeV, both (Off.)\\
 & &  $19^{\mathrm{h}}16^{\mathrm{m}}54^{\mathrm{s}}$ & $-30^\circ25{}^\prime29{}^{\prime\prime}$ & $7.522$ & $-18.420$ & $1.87$ & $0.691$ & 4FGL
\enddata
\tablenotetext{*}{Semi-major axis of the 95\% confidence ellipse for the source localization.}
\tablenotetext{\dagger}{Minimum detectable pulsed fraction, assuming an isolated pulsar with a narrow pulse profile, and no timing noise or glitches (see Section~\ref{s:search}).}
\tablenotetext{\ddagger}{Spectral analysis in which source was identified as a candidate pulsar, and from which photon weights were calculated (see Section~\ref{s:sources}). Where the source is identified as a pulsar candidate from the inner-Galaxy analysis with both ``Officical'' and ``Alternative'' diffuse emission models, the one used to compute the photon weights is given in brackets.}

\tablenotetext{a}{New gamma-ray pulsar discovered in this work.}
\tablenotetext{b}{Radio MSP since discovered in searches of this gamma-ray source.}
\tablenotetext{c}{Associated with an active galaxy in \citetalias{4FGL-DR4}.}
\tablenotetext{d}{Associated with an SNR or pulsar wind nebula in \citetalias{4FGL-DR4}.}
\tablenotetext{e}{Associated with a radio or X-ray source of unknown type.}
\end{deluxetable*}

\section{Dependence of $W^2$ on pulsar photon flux}
\label{a:W2}
The weight for a photon with energy $E$ and arrival direction
$\vec{\Omega}$ is calculated as the ratio of the expected photon flux ($F_{\rm psr}$, in units of photons~cm$^{-2}$~s$^{-1}$~MeV$^{-1}$~steradian$^{-1}$), after
convolution of a point source at the nominal pulsar position 
with the energy-dependent LAT point spread function,
 divided by the total expected flux (i.e., the sum of the pulsar
flux and the total background flux, $F_{\rm psr} + F_{\rm bkg}$),
\begin{equation}
w\left(E,\vec{\Omega}\right) = \frac{F_{\rm psr} \left(E,\vec{\Omega}\right)}{F_{\rm psr}
  \left(E,\vec{\Omega}\right) + F_{\rm bkg} \left(E,\vec{\Omega}\right)}\,,
\end{equation}
Summation over observed photons is equivalent to integrating the total photon
flux multiplied by the accumulated instrument exposure ($\epsilon$, with units of cm$^{2}$~s) over energies
and directions, 
\begin{equation}
    \sum_j 1 \approx \int \int \left(F_{\rm psr}
  \left(E,\vec{\Omega}\right) + F_{\rm bkg} \left(E,\vec{\Omega}\right)\right) \epsilon\left(E,\vec{\Omega}\right) \,dE\,d\vec{\Omega} \,,
\end{equation}
and hence the sum of the squared photon weights is
\begin{equation}
\begin{aligned}
\sum_j w_j^2 &\approx \int \int w^2\left(E,\vec{\Omega}\right) \left(F_{\rm psr}
  \left(E,\vec{\Omega}\right) + F_{\rm bkg} \left(E,\vec{\Omega}\right)\right) \epsilon\left(E,\vec{\Omega}\right) \,dE\,d\vec{\Omega} \,, \\
&\approx \int \int \frac{F_{\rm psr}^2 \left(E,\vec{\Omega}\right)}{F_{\rm psr}
  \left(E,\vec{\Omega}\right) + F_{\rm bkg} \left(E,\vec{\Omega}\right)} \epsilon\left(E,\vec{\Omega}\right) \,dE\,d\vec{\Omega} \,.
\end{aligned}
\end{equation}
In the inner Galaxy, and in the faint-source regime, the gamma-ray background is
high compared to the pulsar flux at all energies, and hence we can ignore the
pulsar flux contribution to the denominator. To see how this sum depends on the
pulsar flux, we extract the overall normalizing factors from the pulsar and
background spectra as $F\left(E,\vec{\Omega}\right) = F_0 f\left(E,\vec{\Omega}\right)$, and the integral becomes
\begin{equation}
\sum_j w_j^2 \approx \frac{F_{\rm psr,0}^2}{F_{\rm bkg,0}} \int \int \frac{f_{\rm psr}^2 \left(E, \vec{\Omega}\right)}{f_{\rm bkg} \left(E,\vec{\Omega}\right)} \epsilon \left(E, \vec{\Omega}\right)\,dE\,d\vec{\Omega}\,,
\end{equation}
The integral depends on the background map, and on the location of the pulsar
and the shape of its spectrum, but for any given point source it is a constant that
we do not require for the purposes of comparing semi-coherent and coherent
search sensitivities. Hence, our pulsation spectral signal-to-noise ratios are
proportional to the square of the pulsar flux.

\end{document}